\documentclass[twocolumn,secnumarabic,amssymb, nobibnotes, superscriptaddress, nofootinbib,aps,prd]{revtex4-2}
\usepackage[colorlinks,citecolor=blue,linkcolor=blue,anchorcolor=blue,filecolor=blue, urlcolor=blue]{hyperref}
\usepackage{natbib}
\usepackage{xcolor}

\usepackage{amsmath}
\usepackage{adjustbox}
\usepackage{graphicx}
\usepackage{booktabs}       
\usepackage{amsfonts}       
\usepackage{nicefrac}       
\usepackage{microtype}
\usepackage{cleveref}
\usepackage{xcolor}
\usepackage{multirow}
\usepackage{makecell}
\usepackage{comment}
\usepackage{todonotes}
\usepackage{array}
\usepackage{orcidlink}
\usepackage{float}
\usepackage{enumerate}
\usepackage[normalem]{ulem}

\newcommand{\ishfaq}[1]{{\color{blue}{#1}}}

\begin{document}

\title{Constraining Axion-Like Particle mediated Dark Matter with Observational Constraints: A Statistical and Machine Learning Approach} 

\author{Prashant Thakur \orcidlink{0000-0003-4189-6176}}
\email{prashantthakur1921@gmail.com}
\affiliation{Department of Physics, Yonsei University, Seoul, 03722, South Korea}

\author{Aravind Taridalu \orcidlink{0000-0003-4189-6176}}
\email{p2021@goa.bits-pilani.ac.in}
\affiliation{Department of Physics, BITS-Pilani, K. K. Birla Goa Campus, Goa 403726, India}

\author{Ishfaq Ahmad Rather~\orcidlink{0000-0001-5930-7179}}
\email{rather@astro.uni-frankfurt.de}
\affiliation{Institut f\"{u}r Theoretische Physik, Goethe Universit\"{a}t, 
Max-von-Laue-Str.~1, D-60438 Frankfurt am Main, Germany}

\author{Tanech Klangburam~\orcidlink{0009-0002-0406-846X}}
\email{klangburam.t@gmail.com}
\affiliation{Khon Kaen Particle Physics and Cosmology Theory Group (KKPaCT)}

\author{Chakrit Pongkitivanichkul~\orcidlink{0000-0003-1478-7732}}
\email{chakpo@kku.ac.th}
\affiliation{Department of Physics, Faculty of Science, Khon Kaen University, 123 Mitraphap Rd.,
Khon Kaen, 40002, Thailand}

\begin{abstract}

We present a comprehensive investigation into the phenomenological consequences of axion-like particle (ALP) mediated dark matter (DM) on the structure of neutron stars (NSs). Our analysis is grounded in a well-established relativistic mean-field framework, featuring non-linear mesonic self-interactions constrained by nuclear physics data and modern astrophysical observations. We systematically explore the DM parameter space, spanning DM particle masses $m_\chi \in [0, 1000] ~\mathrm{GeV}$ and DM fermi momenta $q_f \in [0, 0.06]~\mathrm{GeV}$, by generating a vast ensemble of over 30,000 equations of state (EoSs). This analysis is carried out using two representative hadronic EoSs, a stiff one (EoS1) and a soft one (EoS18), with the explicit inclusion of the crustal EoS to properly account for the low-density regime of NSs. 
Employing a multi-tiered statistical filtering scheme, combining voting, likelihood, and kernel density estimation scores, we apply stringent constraints from a suite of multi-messenger observations, including radio and X-ray pulsars, GW170817, and low-mass compact object HESS J1731-347, revealing that models satisfying the PSR~J0614$-$3329 radius bound inherently comply with the HESS constraints, positioning ALP-mediated DM as a viable candidate for explaining low-mass compact objects while still supporting $2\,M_\odot$ NSs. For the stiff EoS, we obtain a lower bound of $m_\chi \gtrsim 43,\mathrm{GeV}$, with score-weighted posteriors favoring $q_f = 0.034^{+0.020}_{-0.012}$ and a broad allowed DM mass range $m\chi \in [101, 949],\mathrm{GeV}$ (median $\sim 466,\mathrm{GeV}$). In contrast, the soft EoS yields no strict lower bound on $m_\chi$, although scenarios with simultaneously large $m_\chi$ and $q_f$ are strongly disfavored. We developed a high-precision supervised interpolation model using \texttt{AutoGluon} to infer DM parameters from reconstructed NS mass--radius curves, achieving $R^2 > 0.998$. Feature-importance analysis indicates that the DM mass $m_\chi$ is mainly constrained by global shape indicators such as the radius ratio $R_{1.6}/R_{1.4}$, whereas the Fermi momentum $q_f$ is primarily determined by the tidal deformability $\Lambda_{1.4}$.

\end{abstract}
\maketitle

\section{Introduction}
Neutron stars (NSs) are the dense end-products of massive stellar evolution, formed during the violent core-collapse phase of a supernova. These compact objects confine about $1$--$2,M_{\odot}$ within a radius of around $10$--$15$ km. A central open question in NS physics concerns the nature of matter at their ultra-dense cores, where densities can reach $5$--$10$ times the nuclear saturation density ($\rho_0$)~\cite{Lattimer:2000nx,Burrows:1986me}. Although major advances have been made, the composition and behavior of matter under such extreme conditions remain uncertain. To probe the physics of these dense cores, extensive research has focused on the equation of state (EoS), which encapsulates the relationship between pressure, density, and temperature for a given system. The EoS not only determines the internal structure of NSs but also governs their observable properties, most notably mass and radius. Decades of theoretical work have produced a wide array of EoS models, from microscopic many-body calculations to phenomenological frameworks, which attempt to describe the complex interactions between nucleons and potential exotic phases, such as hyperonic matter or deconfined quarks~\cite{Gal:2016boi,Curceanu:2019uph,Tolos:2020aln,Akmal:1998cf, Mueller:1996pm, Danielewicz:2002pu, Baym:2017whm}. The advent of multi-messenger astronomy has revolutionized our ability to test these theories. The measurements of
several pulsars with masses of around $2.0\,M_\odot$ from radio observations of binaries \cite{Demorest:2010bx, Antoniadis:2013pzd, Fonseca:2021wxt, NANOGrav:2019jur} necessitate a strongly repulsive interaction so that the NS can withstand
the strong attraction of gravity.  Gravitational wave signals from binary neutron star mergers, such as GW170817, provide stringent constraints on the tidal deformability~\cite{LIGOScientific:2018cki, LIGOScientific:2017vwq, LIGOScientific:2017ync}, while X-ray pulse-profile modeling from missions like the Neutron Star Interior Composition Explorer (NICER) has yielded unprecedentedly precise mass-radius measurements for several pulsars~\cite{riley2019, Miller:2019cac, Salmi:2024aum, Dittmann:2024mbo, Choudhury:2024xbk, miller2021, riley2021, Vinciguerra:2023qxq, Mauviard:2025dmd}. Together, these observations are beginning to delineate the viable parameter space for the nuclear EoS \cite{Barman:2024zuo, Rather:2020lsg, Christian:2019qer}. 

Beyond the complexities of baryonic matter, the profound gravitational potential of an NS renders it an efficient trap for ambient Dark Matter (DM) particles. Over astrophysical timescales, non-annihilating DM can accumulate in the stellar interior, forming a coexisting component that modifies the star’s structure. While cosmological observations, such as the issue of galaxy rotation curves, firmly establish the presence of DM on large scales~\cite{deMartino:2020gfi,Navarro:1995iw}, compact objects like NSs provide a unique laboratory to probe its properties on much smaller scales. Numerous studies have attempted to constrain the DM fraction in NSs using observational data.
Some suggest that current observations favor small admixtures ($f_\chi \lesssim 5\%$)~\cite{Ivanytskyi:2019wxd,Karkevandi:2021ygv}, while others allow for more generous upper bounds up to $f_\chi \leq 20\%$~\cite{Rutherford:2022xeb}. In contrast, several works argue that much larger fractions remain viable, showing that $2\,M_\odot$ stars can be supported even with 15--70\% DM admixtures~\cite{Ciarcelluti:2010ji,Demorest:2010bx,Lynch_2013,Goldman:2013qla}, and in extreme cases such as XTE J1814--338, DM fractions approaching 90\% have been proposed~\cite{Pitz:2024xvh}.

To investigate the role of DM in NSs, various theoretical models have been proposed, with a prominent class of them falling under the single-fluid formalism, where DM and ordinary matter are treated as a single effective fluid with modified properties due to their interactions. Within this framework, several mechanisms have been explored. One scenario involves non-gravitational interactions between DM and baryons, such as those mediated by the Higgs portal~\cite{Das:2018frc,PhysRevD.96.083004,Dutra:2022mxl}. Another model considers the possibility of neutron decay into DM particles, offering a potential resolution to the neutron lifetime anomaly~\cite{PhysRevLett.121.061801,Shirke:2023ktu,Thakur:2024btu}. A further extension includes dark boson-mediated feeble interactions between DM and hadronic matter, which can subtly influence the NS structure~\cite{Sen:2024yim}. The other approach considers only gravitational interactions between dark matter and ordinary matter. In this case, dark matter and hadronic matter are treated as two distinct components, leading to a two-fluid formalism~\cite{Collier:2022cpr, Miao:2022rqj,Emma:2022xjs, Hong:2024sey, Karkevandi:2021ygv,Ruter:2023uzc,Liu:2023ecz,Ivanytskyi:2019wxd,Buras-Stubbs:2024don,Rutherford:2022xeb,Issifu:2025gsq,Arvikar:2025hej, Hajkarim:2024ecp, Biesdorf:2024dor}. This framework allows each fluid to evolve independently under gravity, enabling the study of not only DM cores embedded within neutron stars but also extended DM halos~\cite{Nelson:2018xtr,Bhattacharya:2023stq, Hajkarim:2024ecp, Barbat:2024yvi} surrounding them. Such configurations can significantly influence the star’s global properties, including mass, radius, and tidal deformability \cite{Biesdorf:2024dor, Barbat:2024yvi}. \citet{Thakur:2023aqm,Shirke:2024ymc} focus on investigating the correlations between DM model parameters and various NS properties within this two-fluid framework. \citet{Issifu:2024htq} were the first to investigate the impact of mirror DM on proto-neutron star evolution within the framework of the two-fluid formalism. Numerous candidates for DM particles, including bosonic DM, axions, sterile neutrinos, and various WIMPs, are discussed in \cite{Bertone:2010zza,Bauer:2017qwy,Calmet:2020pub, Pitz:2023ejc, Pitz:2024xvh}. Studies such as~\cite{Sedrakian:2018kdm,Klangburam:2023vjv} have emphasized the strong theoretical motivation for axion-like particles (ALPs), which are light pseudoscalar particles arising from extensions of the Standard Model and are a subject of intense theoretical and observational study~\cite{Sedrakian:2018kdm,Klangburam:2023vjv,article}. In recent years, several studies have also explored the influence of DM on both radial and non-radial oscillation modes of NSs ~\cite{Thakur:2025zhi,Jimenez:2021nmr,Shirke:2024ymc}. A pioneering study by~\citet{article} proposed that stacking the spectra of active galactic nuclei located behind galaxy clusters provides stronger constraints on ALPs and highlighted that the forthcoming Cherenkov Telescope Array (CTA) Observatory will substantially improve these limits across a wider parameter space.

Building upon prior investigations, this work aims to address these challenges through a significant advancement in both physical modeling and statistical methodology. While a recent study by~\citet{Klangburam:2025rcb} explored ALP-mediated DM within the Quantum Hadrodynamics framework \cite{Serot:1997xg, Walecka:1974qa}, we extend this analysis by employing a more sophisticated non-linear relativistic mean-field (RMF) model. Crucially, our RMF model is not arbitrary but is itself rigorously constrained by nuclear matter properties, state-of-the-art chiral effective field theory ($\chi$EFT) predictions \cite{Hebeler:2013nza, Drischler:2017wtt}, and existing astrophysical data through a comprehensive Bayesian analysis~\cite{Malik:2024nva}. Rather than relying on simple exclusion criteria based on individual observations, we introduce a novel statistical framework that synthesizes the combined constraining power of multiple, disparate astrophysical measurements. Through a multi-tiered scoring and filtering scheme, we move beyond simple exclusion plots to derive robust, probability-weighted constraints on the ALP-mediated DM parameter space, namely the DM particle mass ($m_\chi$) and its Fermi momentum ($q_f$). Furthermore, we incorporate a machine learning framework as a complementary tool to assess the predictive power of different NS observables. This approach quantifies how effectively the geometry of the mass–radius relation encodes the imprint of DM, thereby identifying which population-level properties can best guide future observational strategies for constraining the dark sector.

This paper is structured as follows. In Section~\ref{NS1}, we present the RMF model extended to include ALP dark matter and outline the hadronic EoSs considered. Section~\ref{results} reports the main results, analyzing the influence of ALP DM on the neutron star equation of state. In Section~\ref{astro_const}, we apply observational constraints to narrow down the ALP parameter space and use this range for subsequent statistical analysis. Section~\ref{machinelearning} describes how we infer ALP dark matter parameters using a machine learning approach. Finally, Section~\ref{discus&summary} summarizes our key findings, draws conclusions, and highlights possible directions for future research. Appendix \ref{appendix} discusses the effect of DM parameters on the stellar properties using a soft hadronic EoS.

\section{Methodology}\label{NS1}
\subsection{Non Linear RMF model}

In the RMF model with non-linear mesonic contributions, the EoS for nuclear matter is described by the interaction of the scalar-isoscalar meson $\sigma$, the vector-isoscalar meson $\omega$, and the vector-isovector meson $\varrho$. The Lagrangian density is given by \citep{Fattoyev:2010mx,Dutra:2014qga,Malik:2023mnx}
        \begin{equation}
          \mathcal{L_B}=   \mathcal{L}_N+ \mathcal{L}_M + \mathcal{L}_{NL} +\mathcal{L}_{leptons},
\end{equation} 
where
$$\mathcal{L}_{N} = \bar{\Psi}\Big[\gamma^{\mu}\left(i \partial_{\mu}-g_{\omega} \omega_\mu - \frac{1}{2}g_{\varrho} {\boldsymbol{\tau}} \cdot \boldsymbol{\varrho}_{\mu}\right) - \left(m_N - g_{\sigma} \sigma\right)\Big] \Psi,$$
denotes the Dirac Lagrangian density for the neutron and proton doublet with a bare mass $m_N$ in interaction with mesons $\sigma$, $\omega$ and $\varrho$, where $\Psi$ denotes a Dirac spinor, $\gamma^\mu$ are the Dirac matrices, and $\boldsymbol{\tau}$ symbolizes the Pauli matrices.
The $\mathcal{L}_{M}$ is the Lagrangian density for the mesons, given by
\begin{eqnarray}
\mathcal{L}_{M}  &=& \frac{1}{2}\left[\partial_{\mu} \sigma \partial^{\mu} \sigma-m_{\sigma}^{2} \sigma^{2} \right] - \frac{1}{4} F_{\mu \nu}^{(\omega)} F^{(\omega) \mu \nu} + \frac{1}{2}m_{\omega}^{2} \omega_{\mu} \omega^{\mu}   \nonumber \\
  &-& \frac{1}{4} \boldsymbol{F}_{\mu \nu}^{(\varrho)} \cdot \boldsymbol{F}^{(\varrho) \mu \nu} + \frac{1}{2} m_{\varrho}^{2} \boldsymbol{\varrho}_{\mu} \cdot \boldsymbol{\varrho}^{\mu}, \nonumber
\end{eqnarray}
where $F^{(\omega, \varrho)\mu \nu} = \partial^ \mu A^{(\omega, \varrho)\nu} -\partial^ \nu A^{(\omega, \varrho) \mu}$ are the vector meson  tensors, and the term
\begin{eqnarray}
\mathcal{L}_{NL}&=&-\frac{1}{3} b~m_N~ g_\sigma^3 (\sigma)^{3}-\frac{1}{4} c (g_\sigma \sigma)^{4}+\frac{\xi}{4!} g_{\omega}^4 (\omega_{\mu}\omega^{\mu})^{2}  \nonumber \\
&+&\Lambda_{\omega}g_{\varrho}^{2}\boldsymbol{\varrho}_{\mu} \cdot \boldsymbol{\varrho}^{\mu} g_{\omega}^{2}\omega_{\mu}\omega^{\mu}, \label{Lnl}
\end{eqnarray}
includes the non-linear mesonic terms characterized by the parameters $b$, $c$, $\xi$, and $\Lambda_{\omega}$, which manage saturation properties, the symmetry energy, and  the high-density properties of nuclear matter \cite{Boguta:1977xi, Sumiyoshi1995}. The coefficients $g_i$ represent the couplings between the nucleons and the meson fields $i = \sigma, \omega, \varrho$, which have masses denoted by $m_i$.

Finally, the  leptons are described by the Lagrangian density  
$$\mathcal{L}_{leptons}= \bar{\Psi_l}\Big[\gamma^{\mu}\left(i \partial_{\mu}  
-m_l \right)\Big]\Psi_l,$$ 
with  $\Psi_l~(l= e^-, \mu^-)$  the lepton spinor for electrons and muons; leptons are considered non-interacting.

When solving for the EoS of the full system, the particle fractions are determined by enforcing the standard conditions of beta-equilibrium and charge neutrality. The corresponding chemical potentials for all species ($\mu_i$) arise as the Lagrange multipliers that satisfy these constraints at a given baryon density.

\subsection{ALP Dark Matter}

For the DM sector, we adopt the ALP-mediated DM model introduced in our earlier works \cite{Klangburam:2023vjv, Klangburam:2025rcb}. In this framework, the SM is extended by a Dirac fermion $\chi$ (the DM candidate) and a pseudoscalar ALP $a$. We assume that DM interacts with the nucleon through the ALP mediator. The effective Lagrangian is given by
\begin{align}
\mathcal{L}_{\rm DM} &= \bar{\chi}(i\gamma^\mu\partial_\mu - m_\chi)\chi 
+ \frac{1}{2} \partial_\mu a \partial^\mu a - \frac{1}{2} m_a^2 a^2 \nonumber \\
&\quad + \sum_f \frac{m_\chi}{f_a} C_\chi \bar{\chi} i\gamma^5 \chi a 
+ \sum_f \frac{m_f}{f_a} C_f \bar{\Psi} i\gamma^5 \Psi a  
\end{align}
where $f$ is any SM fermion. $m_f$, $m_\chi$, and $m_a$ are the masses of the fermion, DM, and ALP, respectively. We define the effective ALP couplings as
\begin{equation}
g_{aff} = \frac{m_f}{f_a} C_f \qquad \text{and} \qquad g_{a\chi\chi} = \frac{m_\chi}{f_a} C_\chi, 
\end{equation}
where the ALP-fermion couplings $C_f$ are assumed to be universal for any SM fermion, i.e., $C_f$ and $g_{aff} \propto m_f$. The ALP-mediated DM framework has been investigated across different regimes in the literature. For instance, freeze-in and freeze-out scenarios have been analyzed in~\cite{Bharucha:2022lty,Ghosh:2023tyz,Dror:2023fyd,Armando:2023zwz,Allen:2024ndv}, while astrophysical constraints have been explored in~\cite{Klangburam:2023vjv,Klangburam:2025rcb,Yang:2024jtp}. In the present work, we focus on the scenario where DM interacts with nucleons inside NSs.

The total Lagrangian of the system is
\[
\mathcal{L} = \mathcal{L}_{\text{B}} + \mathcal{L}_{\text{DM}}.
\]

To analyze the system's behavior, we apply the RMF approximation to the total Lagrangian. In this approach, the meson and ALP fields are treated as classical fields, replaced by their mean values. This approximation allows us to derive the coupled equations of motion for the fields and fermions, as well as the expressions for the effective energies, effective masses, and the equation of state of the system.

The total energy density ($\mathcal{E}$) and pressure ($P$) of the system incorporating the contributions from both fermions and the mean fields is given as:

\begin{align}
\mathcal{E} &= \frac{\gamma}{2\pi^2} \int_0^{k_f}dk\ k^2\sqrt{k^2 + \widetilde{m}^2}  + \frac{\gamma_\chi}{2\pi^2} \int_0^{q_f}dq\ q^2\sqrt{q^2 + \widetilde{m}_\chi^2} \nonumber \\
& \quad + \frac{1}{2}m_\sigma^2\sigma_0^2 + \frac{1}{2}m_\omega^2 \omega_0^2 + \frac{1}{2}m_\varrho^2 \varrho_0^2 + \frac{1}{2}m_a^2 a_0^2 \nonumber \\
& \quad + \frac{b\ m_n }{3}\left( g_\sigma \sigma_0 \right)^3 + \frac{c}{4}\left( g_\sigma \sigma_0 \right)^4 + \frac{\xi}{8} \left(  g_\omega\omega_0 \right)^4 \nonumber \\
& \quad + 2\Lambda_\omega \left( g_\omega \omega_0\right)^2 \left( g_\varrho \varrho_0\right)^2 , \\
P &= \frac{\gamma}{6\pi^2} \int_0^{k_f} dk\ \frac{k^4}{\sqrt{k^2 + \widetilde{m}^2}} + \frac{\gamma_\chi}{6\pi^2} \int_0^{q_f} dq\ \frac{q^4}{\sqrt{q^2 + \widetilde{m}_\chi^2}}  \nonumber \\
& \quad - \frac{1}{2}m_\sigma^2\sigma_0^2 + \frac{1}{2}m_\omega^2 \omega_0^2 + \frac{1}{2}m_\varrho^2 \varrho_0^2 - \frac{1}{2}m_a^2 a_0^2 \nonumber \\
& \quad - \frac{b\ m_n }{3}\left( g_\sigma \sigma_0 \right)^3 - \frac{c}{4}\left( g_\sigma \sigma_0 \right)^4 + \frac{\xi}{8} \left(  g_\omega\omega_0 \right)^4 \nonumber \\
& \quad + 2\Lambda_\omega \left( g_\omega \omega_0\right)^2 \left( g_\varrho \varrho_0\right)^2.
\end{align}

Here, $\gamma$ and $\gamma_\chi$ are the spin degeneracy factors for nucleons and DM, respectively. 

In this study, we employ EoS1 and EoS18 from the set of 21 unified NS EoSs presented by~\citet{Malik:2024nva}, which are constrained by nuclear matter properties, state-of-the-art $\chi$EFT calculations for low-density neutron matter, and astrophysical observations. These two representative models were selected to span the range of viable EoS behaviors, from stiff to soft.  

EoS1 represents the stiffest EoS within the dataset, supporting the highest maximum mass of $2.74\,M_\odot$ and exhibiting the largest tidal deformability ($\Lambda_{1.4} = 844$) among all. Its pressure and speed of sound profiles consistently reflect its stiff nature at high densities.

Conversely, EoS18 represents one of the softer EoSs, particularly among those that satisfy all the astrophysical and nuclear constraints imposed in the paper. While not the absolute softest in the entire dataset, its properties, such as a maximum mass of $2.26\,M_\odot$ and a tidal deformability of $\Lambda_{1.4} = 487$, place it on the softer side of the fully compliant EoSs, thus allowing for a comprehensive exploration of the EoS parameter space in our investigation. The key parameters as well as the NS properties for both EoS1 and EoS18 are shown in Table~\ref{tab:EoS1_EoS18_combined}.

Previous studies have shown that, in the regime where axions are trapped inside NSs, the ALP mean field $a_0$ remains negligible. As a result, the ALP parameters have no significant influence on the EoS. We therefore fix their reference values to $m_a = 100~\mathrm{GeV}$, $g_{aff} = 10^{-3}$, and $g_{a\chi\chi} = 10^{-3}$, as done in the previous study \cite{Klangburam:2025rcb}, and restrict our analysis to the DM parameters $m_\chi$ and $q_f$.

\begin{table}[htbp!]
\centering
\caption{Parameters and NS properties for EoS1 and EoS18. The upper section lists the Lagrangian parameters, while the lower section shows the corresponding neutron star properties \cite{Malik:2024nva}.}
\label{tab:EoS1_EoS18_combined}
\renewcommand{\arraystretch}{1.75}
\begin{tabular}{cp{1.75cm}p{1.75cm}}
\hline\hline
Parameter & EoS1 & EoS18 \\
\hline
$g_\sigma$         & 10.411847 & 9.220247 \\
$g_\omega$         & 13.219028 & 11.170082 \\
$g_\rho$           & 11.180337 & 11.087122 \\
$B$                & 2.541001  & 4.036904 \\
$C$                & $-3.586261$ & $-4.553914$ \\
$\xi$              & 0.000845  & 0.003810 \\
$\Lambda_\omega$   & 0.027999  & 0.040742 \\
\hline
$M_{\max}$ [$M_\odot$] & 2.74 & 2.26 \\
$R_{\max}$ [km]        & 13.03 & 11.40 \\
$R_{1.4}$ [km]         & 13.78 & 12.70 \\
$R_{2.08}$ [km]        & 14.04 & 12.31 \\
$\Lambda_{1.4}$        & 844 & 487 \\
$c_s^2$ [$c^2$]        & 0.713 & 0.596 \\
\hline
\end{tabular}
\end{table}

\subsection{Fraction of Dark Matter}\label{fraction}
In the present work, DM and baryonic components are treated as a single thermodynamic fluid interacting through an ALP–mediated coupling that modifies the total energy density and pressure. 
Hence, unlike in the two-fluid formalism, where the DM and baryonic fluids evolve independently and a separate dark-matter gravitational mass $M_\chi$ can be defined, our single-fluid treatment naturally embeds both components into a single equilibrium configuration with a common radius $R$. For completeness and for comparison with recent single-fluid analyses~\cite{Shirke:2023ktu,Das:2025pjl,Shirke:2024ymc}, we estimate the fractional dark-matter content using the ratio of the DM energy density to the total energy density integrated over the stellar volume, defined as
\begin{equation}
f_\chi = \frac{\int_0^R \varepsilon_\chi(r)\, dV}{\int_0^R \varepsilon_{\rm tot}(r)\, dV}
= \frac{M_\chi}{M_{\rm tot}},
\end{equation}
where $\varepsilon_\chi(r)$ and $\varepsilon_{\rm tot}(r)$ are, respectively, the local DM and total energy densities. This quantity provides an approximate measure of the dark-matter fraction within the single-fluid RMF framework.



\label{oscillaltions}


\section{Results and Analysis}

\label{results}
\begin{figure*}[htbp]
  \centering
  \begin{minipage}[b]{0.47\linewidth}
    \centering
    \includegraphics[width=\linewidth]{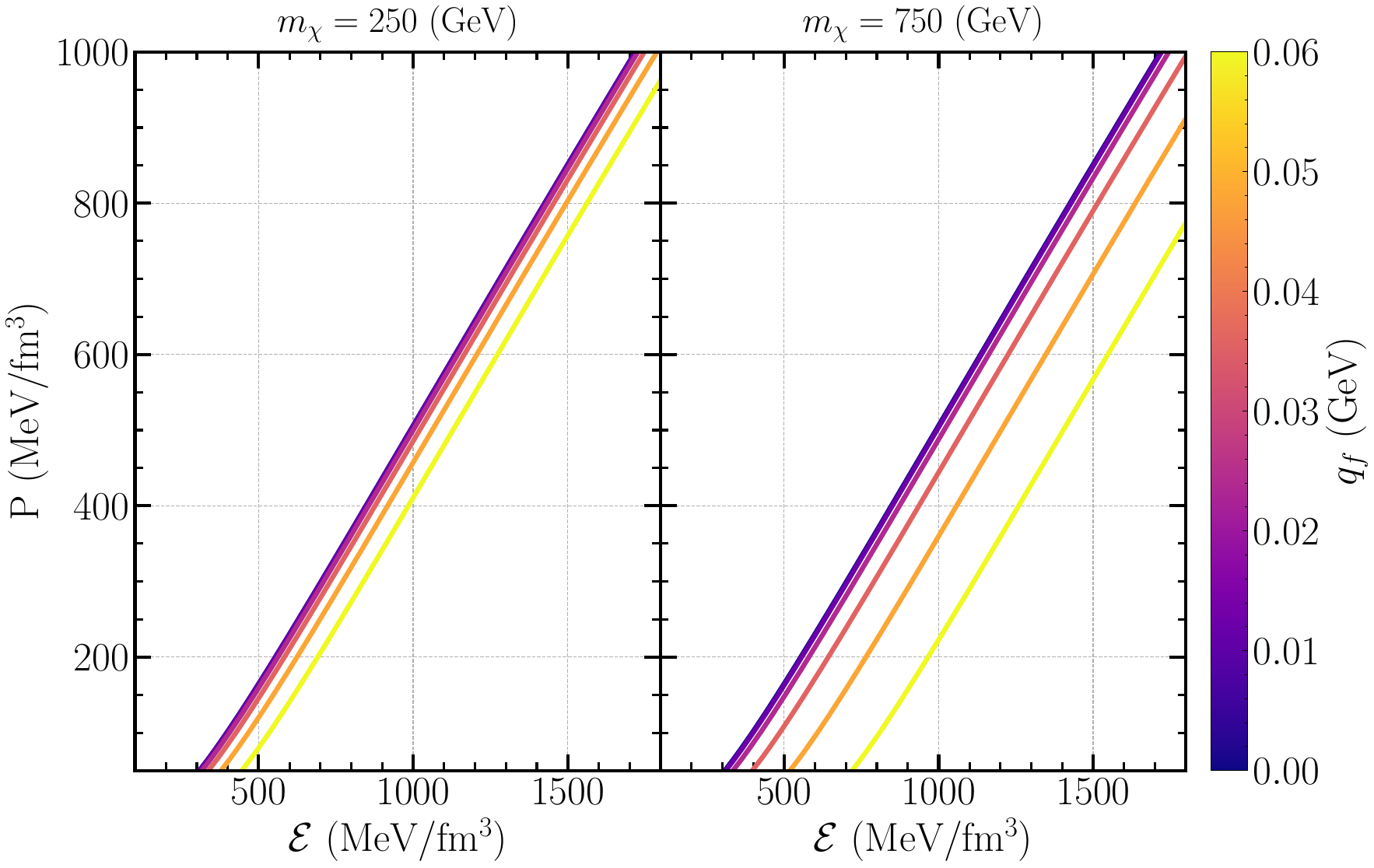}
  \end{minipage}\hspace{0.02\linewidth}%
  \begin{minipage}[b]{0.47\linewidth}
    \centering
    \includegraphics[width=\linewidth]{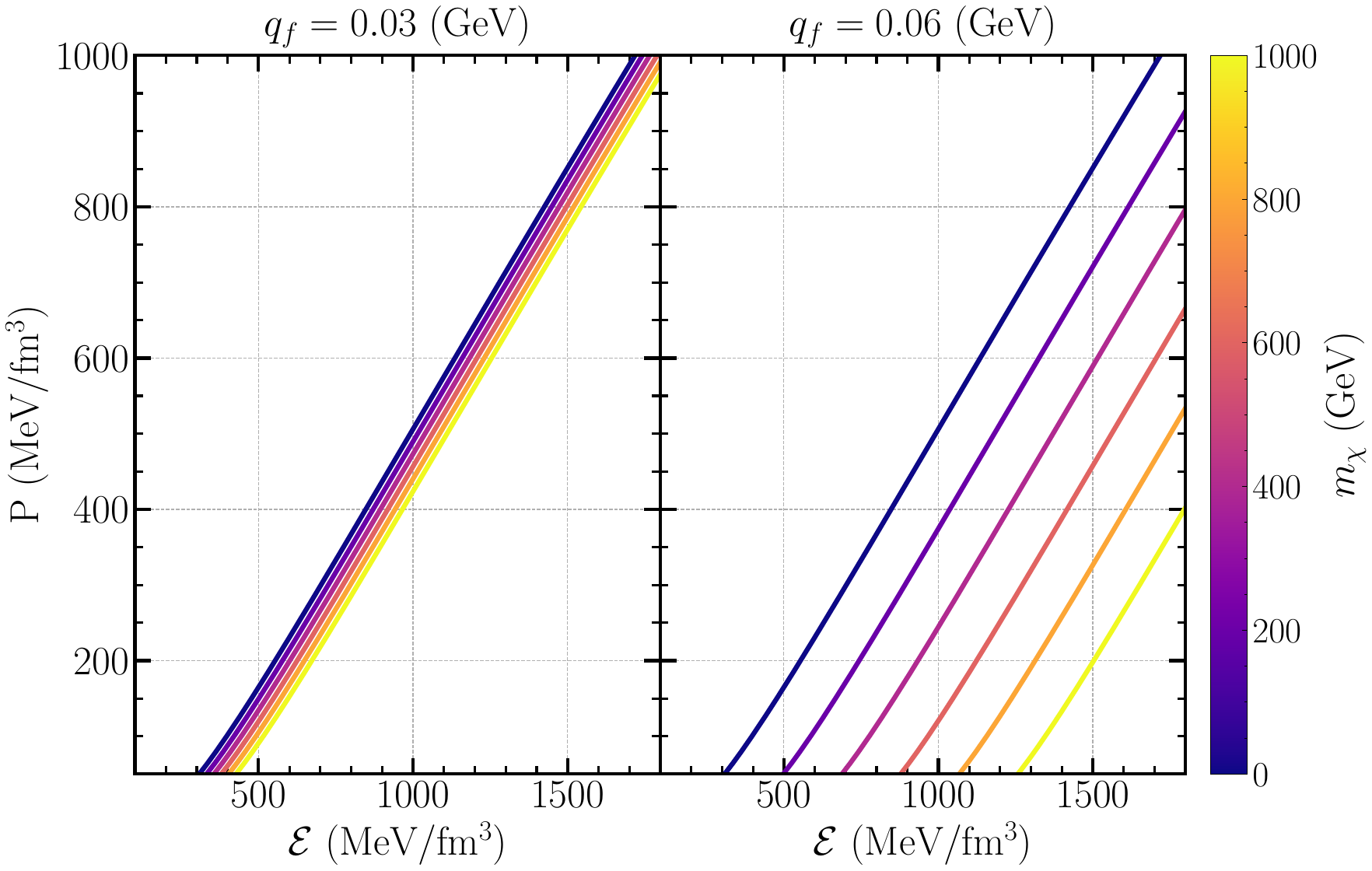}
  \end{minipage}
  \caption{The equation of state (EoS), Pressure ($P$) versus Energy Density ($\mathcal{E}$), for stiff hadronic EoS1. Left: EoS for fixed DM masses ($m_{\chi}$) of 250 and 750~GeV, with varying values of the parameter $q_{f}$. Right: EoS for fixed $q_{f}$ values of 0.03 and 0.06, with varying DM masses $m_{\chi}$. The color bars indicate the value of the varying parameter in each case.}
  \label{EoS_fix}
\end{figure*}

With the modeling framework established, we now turn to the impact of DM parameters on the EoS and the resulting mass–radius MR relation, focusing on the stiff hadronic EoS (EoS1) discussed above.

Figure~\ref{EoS_fix} presents the EoS, showing pressure ($P$) as a function of energy density ($\mathcal{E}$), for a DM model mediated by ALPs. The figure systematically explores the influence of two key parameters: the DM particle mass ($m_{\chi}$) and its Fermi momentum ($q_{f}$). The left panel illustrates the impact of varying $q_{f}$ from 0 to 0.06~GeV, with the DM particle mass held fixed at $m_{\chi} = 250$~GeV (left panel) and $m_{\chi} = 750$~GeV (right panel). With increasing $q_{f}$, the EoS becomes progressively softer, indicating a decrease in pressure at a given energy density. This softening arises from the non-gravitational interactions between DM and baryonic matter, which alter the microphysical properties of the system, such as particle distributions and effective masses, ultimately reducing the pressure support for the same energy density. Conversely, the plot on the right panel illustrates the impact of varying the particle mass $m_{\chi}$ from 0 to 1000~GeV, while the $q_{f}$ value is fixed at 0.03~GeV (left panel) and 0.06~GeV (right panel).
The color bar in this case represents the mass $m_{\chi}$ in GeV. As $m_{\chi}$ increases, the EoS becomes softer. This trend is more prominent at higher $q_{f}$, indicating a stronger sensitivity of the EoS to $m_{\chi}$ when more DM is present in the system.  These plots display that increasing either the Fermi momentum or DM particle mass, while fixing the other one, leads to a softer EoS, and this softness enhances at higher values of $q_{f}$ and $m_{\chi}$.

 The inclusion of crustal effects is essential for a realistic description of NS structure. In this work, we explicitly implement this refinement by incorporating a crustal EoS to account for the low-density regime, which was not considered in the previous study \cite{Klangburam:2025rcb}. The outer crust is modeled using the well-established Baym-PethickSutherland (BPS EoS) over the density range $\rho \sim 4.8 \times 10^{-9}~\mathrm{fm}^{-3}$ to $2.6 \times 10^{-4}~\mathrm{fm}^{-3}$~\cite{Baym:1971pw,Negele:1971vb}. To connect the outer crust with the high-density core EoS, we introduce an intermediate inner crust region described by a polytropic relation of the form $P(\mathcal{E}) = a_1 + a_2 \mathcal{E}^{\gamma}$. The parameters $a_1$ and $a_2$ are fixed by requiring smooth continuity of both pressure and energy density at the boundaries with the outer crust and the core. In this study, we adopt a polytropic index $\gamma = 4/3$, a choice motivated by its ability to capture the stiffness of matter in the inner crust, consistent with prior works~\cite{Malik:2017cdq}. The matching procedure ensures thermodynamic consistency by maintaining the continuity of pressure and chemical potential across the crust–core transition~\cite{rather2020effect,Fortin:2016hny}.

\begin{figure*}[htbp]
  \centering
  \begin{minipage}[b]{0.47\linewidth}
    \centering
    \includegraphics[width=\linewidth]{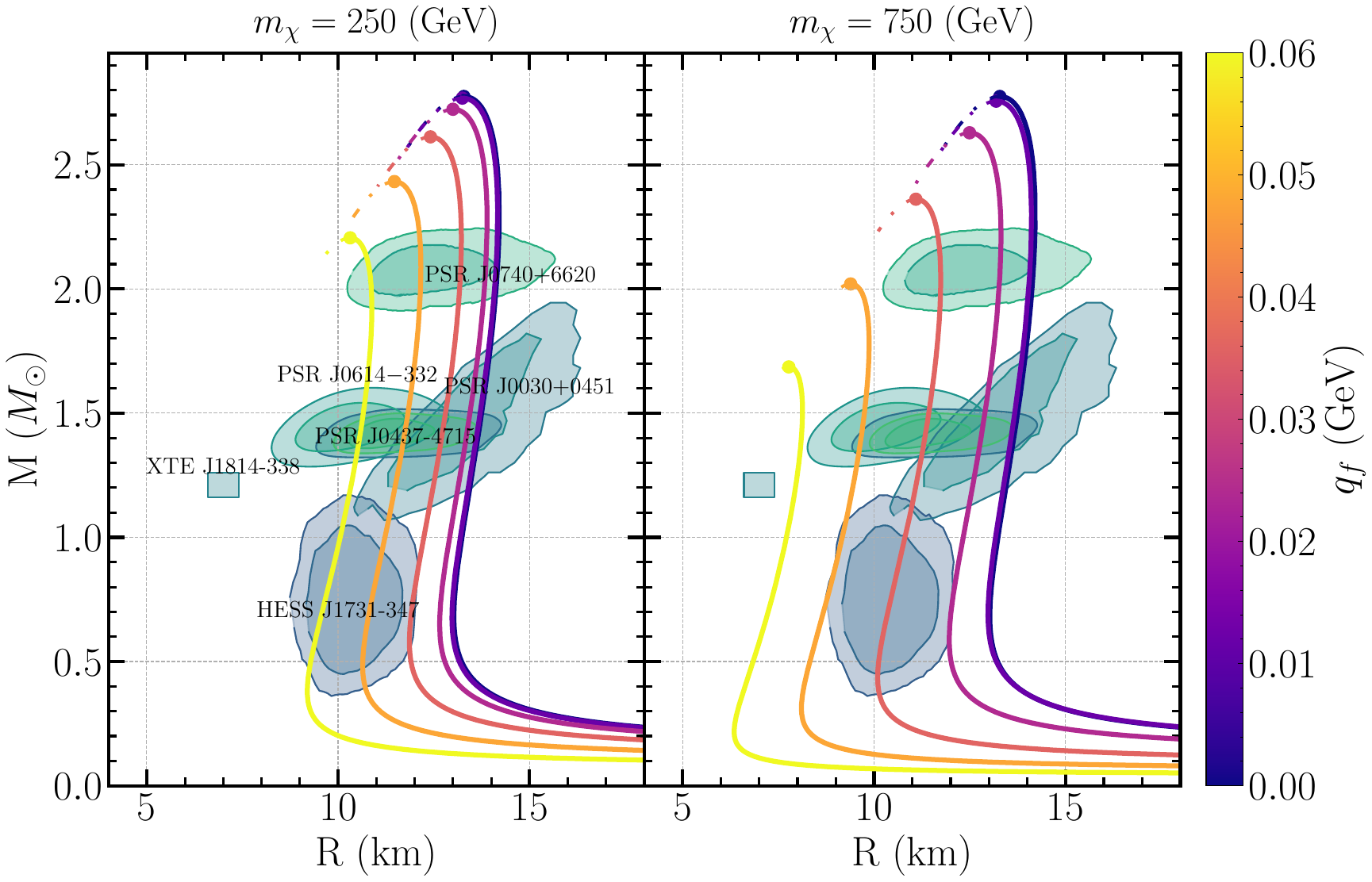}
  \end{minipage}\hspace{0.02\linewidth}%
  \begin{minipage}[b]{0.47\linewidth}
    \centering
    \includegraphics[width=\linewidth]{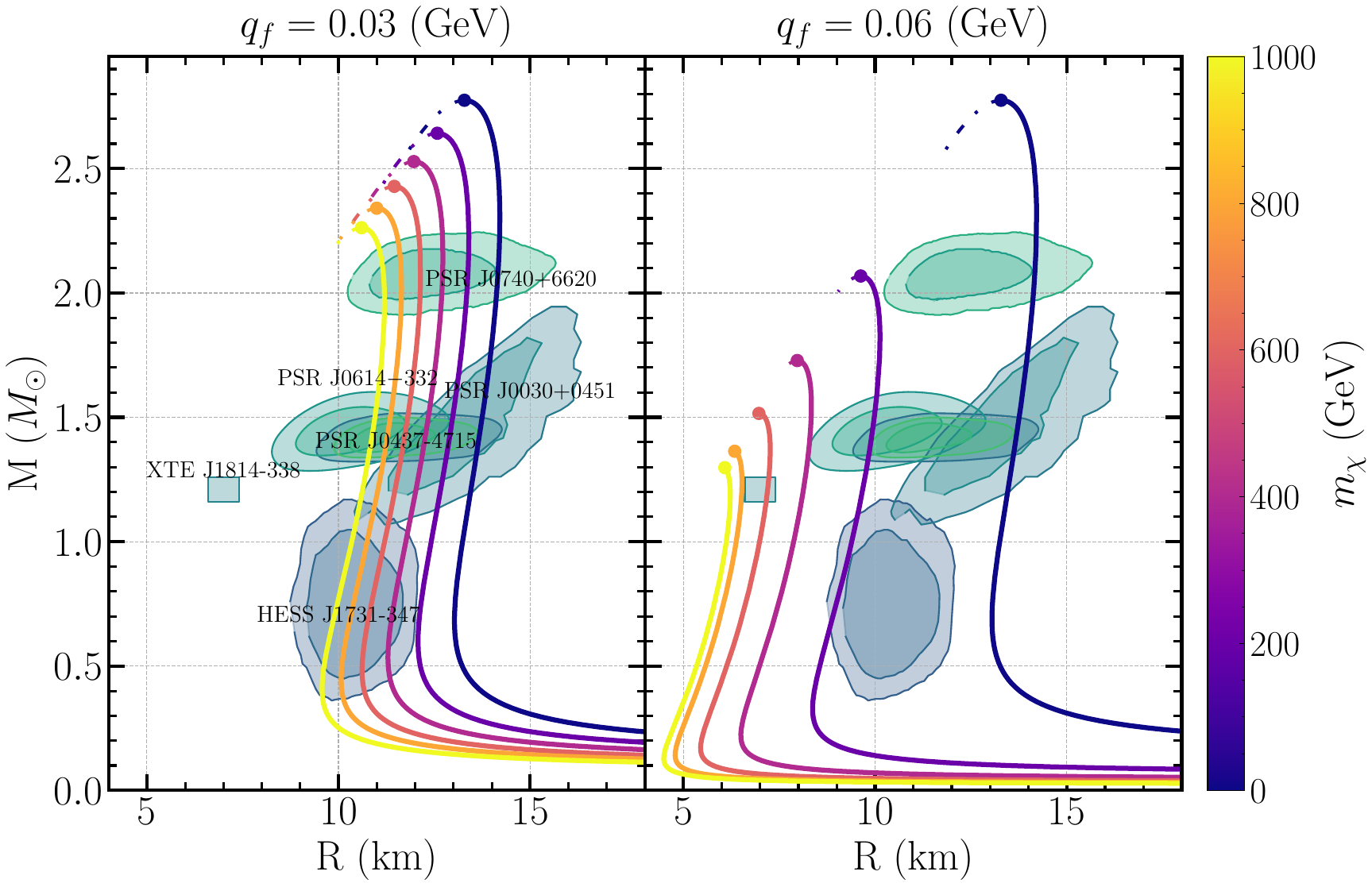}
  \end{minipage}
  \caption{Mass-radius (MR) relations for NSs admixed with DM. highlighting the influence of DM particle mass $m_{\chi}$ and Fermi momentum $q_{f}$, for stiff hadronic EoS1. 
Left: MR for fixed $m_{\chi}$ at 250 GeV (left panel) and 750 GeV (right panel) with $q_{f}$ varying in the range $[0, 0.06]$ GeV. 
Right: MR for fixed $q_{f}$ at 0.03 GeV (left panel) and 0.06 GeV (right panel), with $m_{\chi}$ varying in the range $[0, 1000]$ GeV.  The solid lines represent the stable configurations up to the maximum mass, denoted by a solid dot. The dashed line represents the unstable configurations.
Colorbars denote the respective varying parameters. The plots are overlaid with observational constraints, corresponding to the 68\% and 95\% confidence intervals for PSR J0030+0451~\cite{riley2021, Miller:2019cac}, PSR J0740+6620~\cite{riley2021, miller2021}, PSR J0614$-$3329~\cite{Mauviard:2025dmd}, and PSR J0437-4715~\cite{Choudhury:2024xbk} from NICER measurements. Additional constraints include XTE J1814$-$338~\cite{Kini:2024ggu} and HESS J1731$-$347~\cite{2022NatAs...6.1444D}.}
\label{MR_fix}
\end{figure*}

Figure~\ref{MR_fix} illustrates the impact of DM parameters on the mass--radius relations of DM-admixed NS models using the stiff hadronic EoS1. The left plot corresponds to the MR relation with varying DM Fermi momentum $q_f$ at fixed DM particle masses, $m_\chi = 250~\mathrm{GeV}$ (left panel) and  750~GeV (right panel). Each curve corresponds to a different value of $q_f$, ranging from $0~\mathrm{GeV}$ (no DM) to $0.06~\mathrm{GeV}$, with the colorbar on the right indicating the value of $q_f$ associated with each curve. Increasing the $q_f$ value leads to an increase in the DM content within the star, resulting in stronger gravitational effects and a softer effective EoS. As a result, the MR relation systematically shifts toward lower maximum masses and smaller radii. This trend is visible in both panels: for a fixed $m_\chi$, the stars become more compact and less massive as more DM is admixed. Furthermore, increasing the dark matter mass from 250 to 750~GeV enhances this suppression: the shift in the MR is more towards low values, as shown in the right panel. This is because heavier DM particles contribute more significantly to the star’s gravitational potential for the same Fermi momentum, amplifying the effects. Overlaid on the plot are key observational constraints shown as shaded regions: corresponding to the 68\% and 95\% confidence intervals for several well-studied pulsars: PSR J0030+0451~\cite{riley2019, Miller:2019cac} and PSR J0740+6620~\cite{riley2021, miller2021, Salmi:2024aum, Dittmann:2024mbo}, PSR J0614$-$3329~\cite{Mauviard:2025dmd} from NICER X-ray data. Additionally, the latest NICER constraint for PSR J0437--4715 is shown~\cite{Choudhury:2024xbk}. Additional constraints include for XTE J1814$-$338~\cite{Kini:2024ggu} and the low-mass compact object HESS J1731$-$347 as reported by~\cite{2022NatAs...6.1444D}. These provide direct observational boundaries on viable NS models. Notably, for higher values of $q_f$, the predicted MR relations deviate more significantly from the observed regions, indicating that large DM admixture is incompatible with current observational constraints. 

The right plot shows models with fixed DM fractions $q_f = 0.03$ (left panel) and 0.06~GeV (right panel), respectively, while varying the dark matter particle mass $m_\chi$ between 0-1000~GeV, as indicated by the colorbar on the right. As $m_\chi$ increases, the MR lines shift progressively toward smaller radii and lower maximum masses, demonstrating the enhanced gravitational influence of heavier DM particles even at fixed DM Fermi momentum. This behavior arises because more massive DM particles carry greater energy density per particle, deepening the gravitational potential and destabilizing the star more efficiently. In particular, for large values of both $q_f$ and $m_\chi$, the gravitational softening becomes so extreme that the maximum mass of the star drops below  $1.4\,M_\odot$, a threshold generally considered to be the lower bound for observed neutron star masses. This suppression indicates that such heavily dark matter-admixed stars cannot support the mass of canonical pulsars, but interestingly fall within the constraints for low-mass compact objects like XTE J1814$-$338 \cite{Kini:2024ggu}. 

Thus, while these extreme configurations are incompatible with most known pulsars, they remain viable candidates for explaining anomalously low-mass and small-radius compact stars. Therefore, Figure~\ref{MR_fix} encapsulates how the presence of DM, controlled via $q_f$ and $m_\chi$, can significantly modify NS structure and provides a framework for constraining DM properties using astrophysical observations. We can compare the effect of including the crust EoS with earlier studies such as Ref.~\cite{Klangburam:2025rcb}, which did not account for crustal contributions. At low densities, the absence of a crust leads to backbending in the mass--radius relation of low-mass NSs (similar to self-bound configurations), as evident in that reference. In contrast, incorporating the crust EoS shifts the MR relations toward slightly larger radii and stabilizes the low-mass branch (gravitationally bound). Although the overall qualitative behavior with varying $q_f$ and $m_\chi$ remains unaffected, the inclusion of the crust yields a more physically accurate description of the outer stellar structure.

All the results presented in Figures~\ref{EoS_fix} and \ref{MR_fix} are based on the hadronic model EoS1, which yields a very stiff EoS and, consequently, a large maximum mass. Although the corresponding dimensionless tidal deformability exceeds the upper limit set by the GW170817 BNS event \cite{LIGOScientific:2018cki}, EoS1 serves as a useful baseline for examining the impact of DM parameters. For comparison, we also consider the softer hadronic model EoS18, which satisfies all current astrophysical constraints, to assess how the viable parameter space evolves with changes in the underlying hadronic EoS. The MR relation for the soft hadronic EoS, EoS18, is discussed in the Appendix \ref{appendix}, for comparison.

\subsection{Dark-Matter Fraction and Stellar Properties}
To demonstrate the relative contribution of DM in the ALP-mediated single-fluid framework, 
we evaluated the DM fraction $f_\chi$ for a representative case with 
$m_\chi = 200~\mathrm{GeV}$ and $m_\chi = 1000~\mathrm{GeV}$. 
The fractions were computed for the maximum-mass configurations following the definition introduced in Section~\ref{fraction}, 
and the results are summarized in Table~\ref{tab:DMfraction_combined} for stiff EoS. For $m_\chi = 200~\mathrm{GeV}$, $f_\chi$ increases from about $0.2\%$ to $25\%$ as $q_f$ rises from $0.01$ to $0.06$, 
whereas for $m_\chi = 1000~\mathrm{GeV}$ the corresponding fraction extends up to approximately $55\%$. 
These trends are broadly consistent with other single-fluid RMF-based analyses~\cite{Shirke:2024ymc}, 
where the authors reported a maximum allowed fraction of $\sim37.9\%$, 
while astrophysically motivated limits yield $f_\chi \lesssim 13.7\%$. 
It should be emphasized, however, that the quantitative value of $f_\chi$ is strongly model-dependent. 
For instance, in Ref.~\cite{Shirke:2024ymc} the DM component arises from the neutron-decay anomaly, 
whereas in our case it results from an ALP-mediated coupling. Under the $2\,M_\odot$ observational constraint~\cite{riley2021, miller2021}, the maximum allowed fractions in our framework are 
$f_\chi \approx 25\%$ for $m_\chi = 200~\mathrm{GeV}$ and $f_\chi \approx 18\%$ for $m_\chi = 1000~\mathrm{GeV}$. For the softer EoS as shown in \ref{tab:DMfraction_combined}, the DM fraction $f_\chi$ exhibits a gradual increase from approximately $0.2\%$ to $21.6\%$ for $m_\chi = 200~\mathrm{GeV}$, and up to $50.5\%$ for $m_\chi = 1000~\mathrm{GeV}$. When the $2\,M_\odot$ observational constraint is imposed, only configurations with $q_f \leq 0.04$ ($f_\chi \lesssim 8.4\%$) for $m_\chi = 200~\mathrm{GeV}$ and $q_f \leq 0.03$ ($f_\chi \lesssim 15.4\%$) satisfy the required mass threshold, thereby representing the astrophysically admissible parameter space. Since the total mass $M(R)$ in this approach already accounts for both baryonic and dark-matter energy densities, the $f_\chi$ values serve as indicative measures rather than representing an independently evolving dark-matter component. A full decomposition of baryonic and dark-matter gravitational masses will be addressed in our forthcoming two-fluid TOV investigation.

\begin{table*}[t!]
\centering
\setlength{\tabcolsep}{7pt} 
\renewcommand{\arraystretch}{1.0}
\caption{Estimated dark-matter fraction ($f_\chi$) and gravitational mass ($M_\chi$) for the maximum-mass configurations corresponding to $m_\chi = 200$ and $1000~\mathrm{GeV}$ 
for stiff and soft EoS. 
The pure-baryonic maximum masses are 
$M_{\mathrm{max}}^{\mathrm{pure}} = 2.745\,M_\odot$ (stiff) 
and $2.263\,M_\odot$ (soft).}
\label{tab:DMfraction_combined}
\begin{tabular}{cccc|cccc}
\hline
\multicolumn{4}{c}{\textbf{Stiff EoS}} & \multicolumn{4}{c}{\textbf{Soft EoS}} \\
\hline
\multicolumn{4}{c}{$m_\chi = 200~\mathrm{GeV}$} & \multicolumn{4}{c}{$m_\chi = 200~\mathrm{GeV}$} \\
\hline
$q_f$ & $M_{\mathrm{max}}^{\mathrm{mixed}}$ [$M_\odot$] & $f_\chi$ [\%] & $M_\chi$ [$M_\odot$] &
$q_f$ & $M_{\mathrm{max}}^{\mathrm{mixed}}$ [$M_\odot$] & $f_\chi$ [\%] & $M_\chi$ [$M_\odot$] \\
\hline
0.01 & 2.73 & 0.19 & 0.0053 & 0.01 & 2.26 & 0.15 & 0.0034 \\
0.02 & 2.70 & 1.50 & 0.0409 & 0.02 & 2.23 & 1.20 & 0.0268 \\
0.03 & 2.64 & 4.78 & 0.1262 & 0.03 & 2.17 & 3.86 & 0.0839 \\
0.04 & 2.48 & 10.27 & 0.2556 & 0.04 & 2.07 & 8.40 & 0.1741 \\
0.05 & 2.29 & 17.46 & 0.3999 & 0.05 & 1.93 & 14.53 & 0.2812 \\
0.06 & 2.06 & 25.46 & 0.5266 & 0.06 & 1.77 & 21.62 & 0.3836 \\
\hline
\multicolumn{4}{c}{$m_\chi = 1000~\mathrm{GeV}$} & \multicolumn{4}{c}{$m_\chi = 1000~\mathrm{GeV}$} \\
\hline
$q_f$ & $M_{\mathrm{max}}^{\mathrm{mixed}}$ [$M_\odot$] & $f_\chi$ [\%] & $M_\chi$ [$M_\odot$] &
$q_f$ & $M_{\mathrm{max}}^{\mathrm{mixed}}$ [$M_\odot$] & $f_\chi$ [\%] & $M_\chi$ [$M_\odot$] \\
\hline
0.01 & 2.71 & 1.10 & 0.0302 & 0.01 & 2.24 & 0.90 & 0.0202 \\
0.02 & 2.58 & 6.73 & 0.1741 & 0.02 & 2.14 & 5.42 & 0.1161 \\
0.03 & 2.26 & 18.47 & 0.4178 & 0.03 & 1.91 & 15.41 & 0.2951 \\
0.04 & 1.88 & 32.23 & 0.6061 & 0.04 & 1.63 & 27.83 & 0.4547 \\
0.05 & 1.53 & 44.69 & 0.6859 & 0.05 & 1.36 & 39.78 & 0.5423 \\
0.06 & 1.24 & 55.23 & 0.6861 & 0.06 & 1.12 & 50.49 & 0.5659 \\
\hline
\end{tabular}
\end{table*}

\section{Astrophysical constraints on the DM parameter space}
\label{astro_const}
In this section, we impose initial constraints on the dark matter parameters, $q_f$ and $m_\chi$. To this end, we evaluate the EoS of the ALP-mediated DM model over the full parameter space. Specifically, $m_\chi$ is varied from 0 to 1000\,GeV in 1\,GeV steps, and $q_f$ from 0.000 to 0.060\,GeV in 0.002\,GeV steps, yielding 30,601 $(m_\chi, q_f)$ pairs. Each pair is injected into the original EoS-generating script using dynamic in-memory patching, which is executed independently to produce the corresponding energy density, pressure, and baryon density arrays. The resulting EoSs are then passed to a TOV solver to compute neutron star observables---mass, radius, and tidal deformability---enabling a systematic analysis of how $m_\chi$ and $q_f$ affect macroscopic NS properties.

The primary constraint on the coupling parameters arises from the requirement that the dense matter EoS must support NSs with masses up to $2\,M_{\odot}$, in line with observations of massive pulsars such as J1614$-$2230 ($M = 1.97 \pm 0.04~M_\odot$)~\cite{Demorest:2010bx}, PSR J0348$+$0432 ($M = 2.01 \pm 0.04~M_\odot$)~\cite{Antoniadis:2013pzd}, and PSR J0740$+$6620 ($M = 2.08 \pm 0.07~M_\odot$)~\cite{Fonseca:2021wxt}. These measurements imply that the EoS must remain sufficiently stiff at high densities.

X-ray observations of PSR J0740$+$6620 using the NICER and X-ray Multi-Mirror Mission (XMM) have yielded radius estimates of $R = 12.39^{+1.30}_{-0.98}$ km~\cite{riley2021} and $R = 13.7^{+2.6}_{-1.5}$ km~\cite{miller2021}. More recent analyses refined these values to $R = 12.49^{+1.28}_{-0.88}$ km~\cite{Salmi:2024aum} and $R = 12.76^{+1.49}_{-1.02}$ km~\cite{Dittmann:2024mbo}. For PSR J0030$+$0451 at 1.4\,$M_{\odot}$, the reported radii are $R = 13.02^{+1.24}_{-1.06}$ km~\cite{Miller:2019cac} and $R = 12.71^{+1.14}_{-1.19}$ km~\cite{riley2019}, in agreement  with the updated constraint of~\citet{Vinciguerra:2023qxq}. The mass-radius measurement for PSR J0437$-$4715 gives $R = 11.36^{+0.95}_{-0.63}$ km for a mass of $M = 1.418 \pm 0.037~M_{\odot}$~\cite{Choudhury:2024xbk}, pointing towards a relatively soft dense-matter EoS. A recent measurement of radius has been inferred for PSR J0614-3329, $R = 10.29^{+1.01}_{-0.86}$ km at $M = 1.44^{+0.06}_{-0.07}~M_{\odot}$ suggesting an even more softer EoS \cite{Mauviard:2025dmd}.

Beyond mass and radius constraints, gravitational wave observations from the GW170817 event~\cite{PhysRevLett.119.161101, PhysRevLett.121.161101} provide a key measurement of the dimensionless tidal deformability at 1.4\,$M_{\odot}$, estimated to be $\Lambda = 190^{+390}_{-120}$~\cite{PhysRevLett.121.161101}.

\begin{figure*}[t]
  \centering
  \begin{minipage}[b]{0.47\linewidth}
    \centering
    \includegraphics[width=\linewidth]{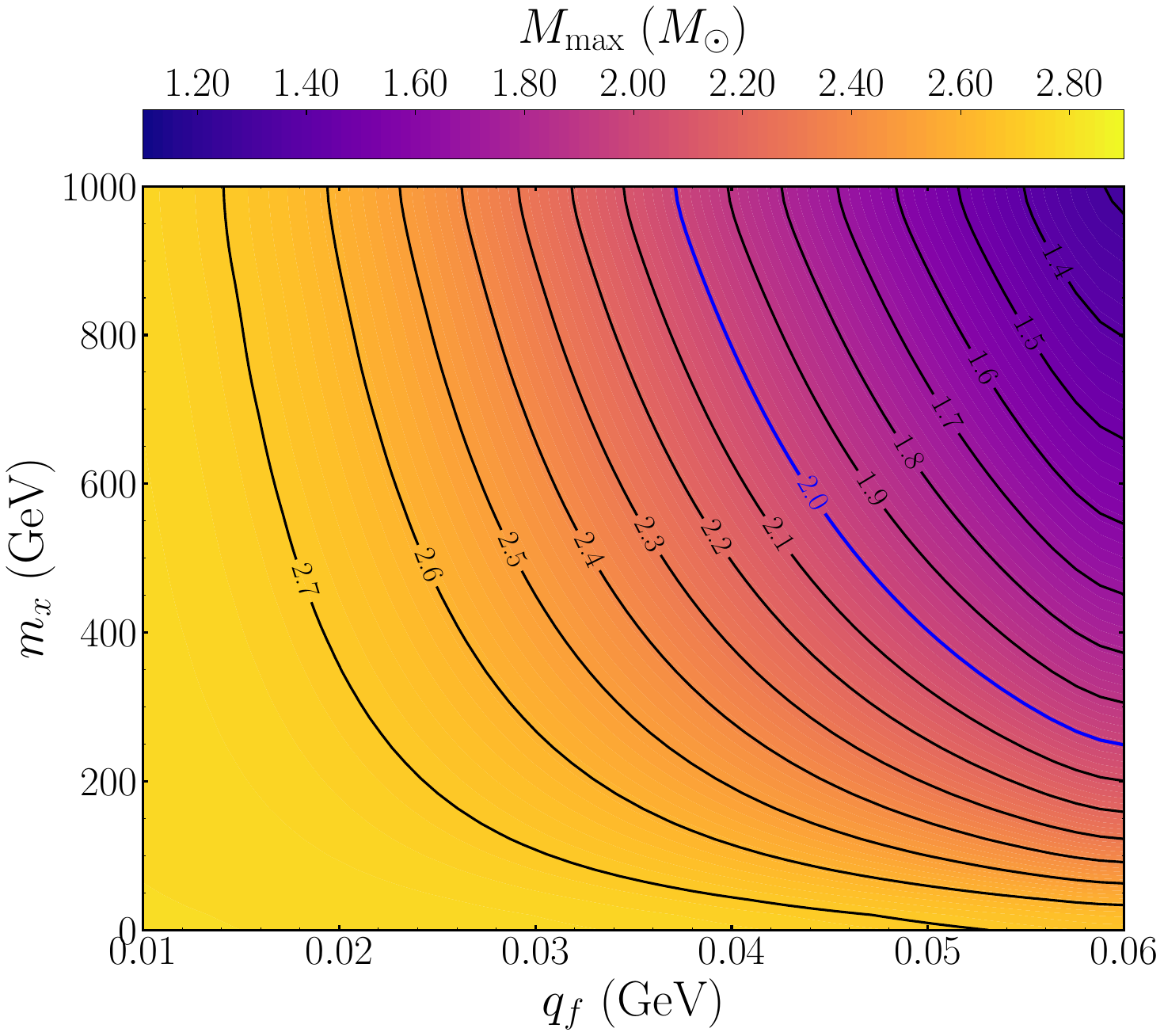}
  \end{minipage}\hspace{0.02\linewidth}%
  \begin{minipage}[b]{0.47\linewidth}
    \centering
    \includegraphics[width=\linewidth]{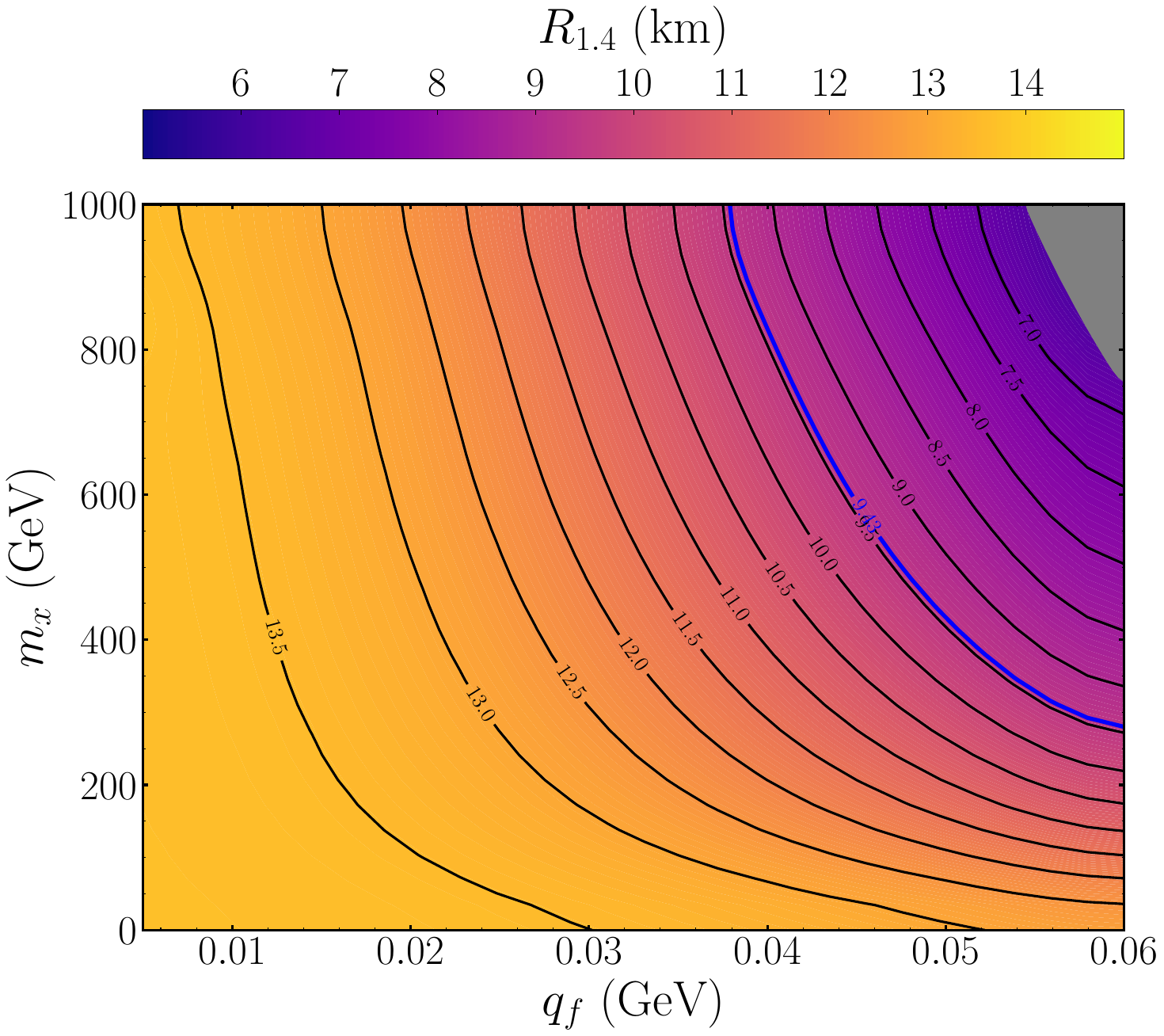}
  \end{minipage}
  \caption{Contour plots in the $q_f$ - $m_\chi$ plane showing (left) the maximum mass of the star and (right) the radius at $1.4\,M_\odot$, for dark matter admixed stars based on the stiff hadronic EoS model EoS1. Contours of the same maximum mass and
$R_{1.4}$ are also shown, respectively. In the left panel, the blue contour marks the $2\,M_\odot$ observational lower limit on the maximum mass \cite{Demorest:2010bx, Antoniadis:2013pzd, NANOGrav:2019jur, Fonseca:2021wxt}. In the right panel, the blue contour corresponds to the NICER measurement of the radius at $1.4\,M_\odot$ \cite{Mauviard:2025dmd}, providing astrophysical constraints on the dark matter parameter space.}
  \label{contour_stiff}
\end{figure*}

\begin{figure*}[htbp]
  \centering
  \begin{minipage}[b]{0.47\linewidth}
    \centering
    \includegraphics[width=\linewidth]{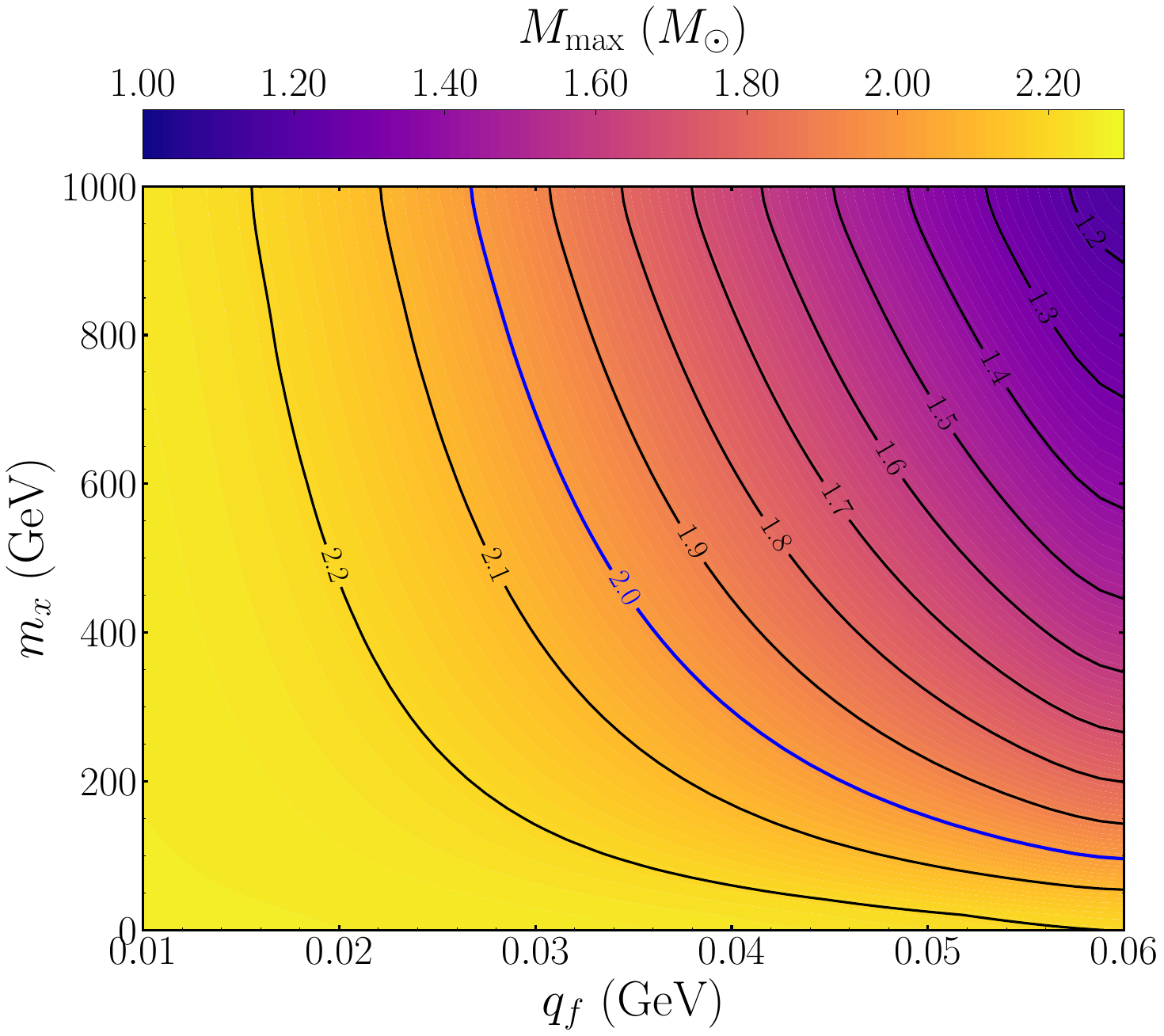}
  \end{minipage}\hspace{0.02\linewidth}%
  \begin{minipage}[b]{0.47\linewidth}
    \centering
    \includegraphics[width=\linewidth]{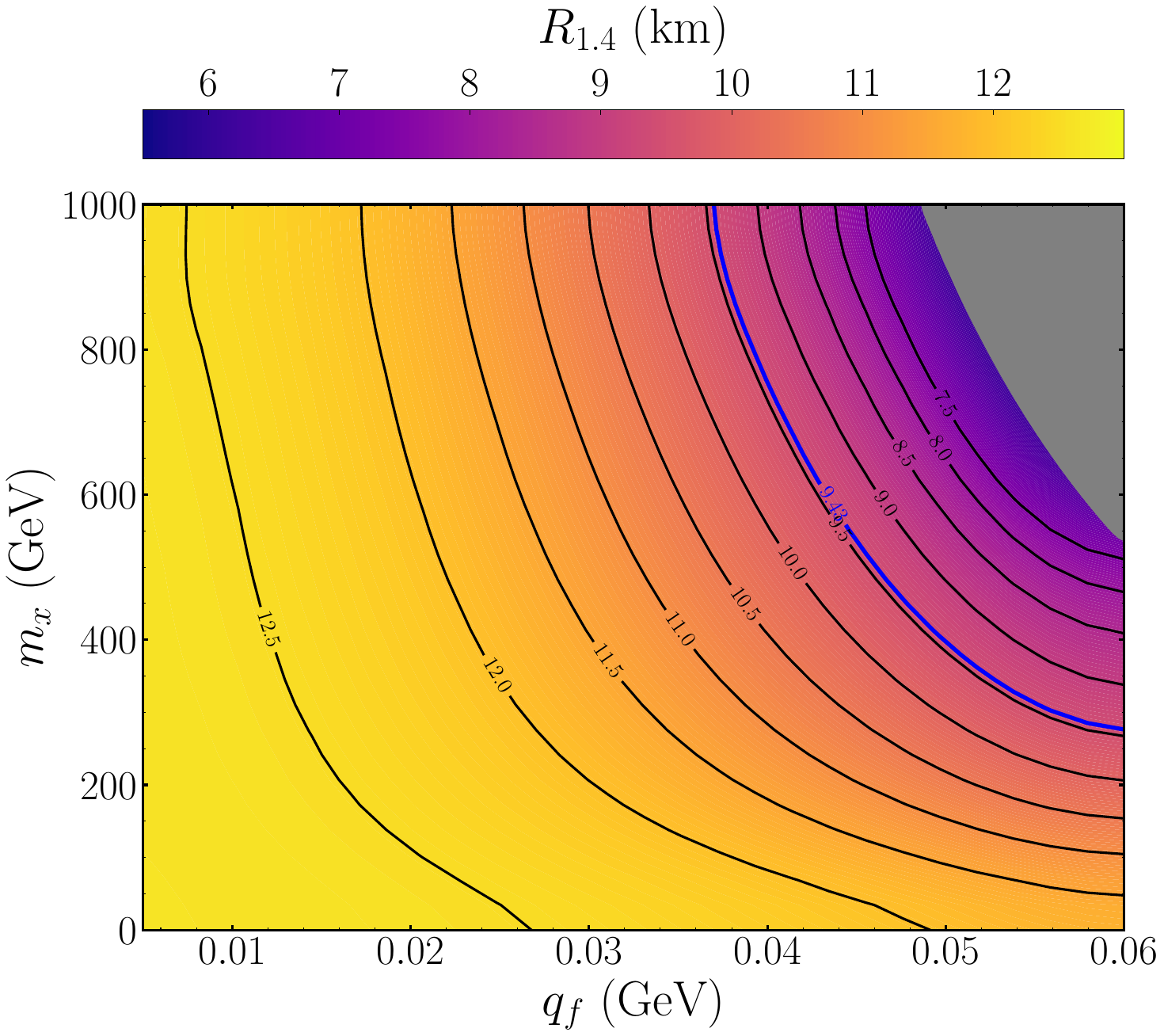}
  \end{minipage}
  \caption{Same as Figure \ref{contour_stiff}, but with a soft hadronic EoS model, EoS18.}
  \label{contour_soft}
\end{figure*}

To ensure continuity and smoothness in our numerical results, particularly for generating contour plots and mass-radius curves, we performed interpolation across discrete data points. Analytical functions were then fitted to key quantities of interest, such as the maximum mass and radius at a fixed stellar mass. This strategy allowed us to obtain smooth contours over the entire parameter space.

Figure~\ref{contour_stiff} represents the contour plot in the $q_f$ - $m_\chi$ plane showing the maximum mass (left plot) and the radius at $1.4\,M_\odot$ (right plot) for the ALP-mediated DM model using the stiff hadronic EoS model, EoS1. The color bar indicates the magnitude of each quantity, with darker (lighter) colors indicating lower (higher) values. Contours of the same maximum mass and $R_{1.4}$ are also shown, respectively. The blue contour line, in the left plot, marks the $2\,M_\odot$ observational lower limit on the maximum mass \cite{Demorest:2010bx, Antoniadis:2013pzd, NANOGrav:2019jur, Fonseca:2021wxt}. From the left plot, we see that at low values of $q_f$, the maximum mass over the whole $m_\chi$ range remains almost the same, $2.7\,M_\odot$, corresponding to the pure hadronic EoS. 
As seen in Figure \ref{contour_stiff}, increasing $q_f$ or $m_\chi$ softens the EoS, which systematically lowers the maximum supported mass ($M_{max}$) of the star. Combinations of high $q_f$ and $m_\chi$ (the upper-right region of the plot) produce an EoS so soft that its $M_{max}$ is below the $2.0~M_\odot$ observational threshold. These EoS models are therefore ruled out \textit{ab initio}, as they cannot account for the existence of massive pulsars \cite{Demorest:2010bx, Antoniadis:2013pzd, NANOGrav:2019jur, Fonseca:2021wxt}. 

The right plot from Figure~\ref{contour_stiff} represents the variation of the radius at $1.4\,M_\odot$ across the $q_f$, $m_\chi$ parameter space studied.  The blue contour line 9.43 km indicates the lower bound on the radius at $1.4\,M_\odot$ based on the recent measurement of PSR J0614$-$3329~\cite{Mauviard:2025dmd}. The corresponding upper bound of 14.25 km, inferred from PSR J0740$+$6620~\cite{Dittmann:2024mbo}, lies beyond the explored parameter space, as the pure hadronic EoS used here yields a radius of 13.78 km.

Figure~\ref{contour_soft} shows the same contour plots in the $q_f$ - $m_\chi$ plane for the ALP DM model but with the soft hadronic EoS model, EoS18. The left panel shows the maximum mass, $M_{max}$, which for a pure baryonic star ($q_f \to 0$) is approximately $2.26\,M_\odot$. As the DM fraction increases with the coupling parameter $q_f$, the additional gravitational load from the DM core softens the overall EoS, leading to a systematic reduction in $M_{max}$. The blue contour, representing the established observational lower limit of $2\,M_\odot$, consequently excludes a significant portion of the parameter space where the star's stability is compromised by the DM admixture. The right panel displays the radius for a canonical $1.4\,M_\odot$ star, $R_{1.4}$. The pure EoS18 yields a relatively compact star with $R_{1.4} \approx 12.70$ km. An increasing DM content further compresses the star, causing its radius to shrink. The blue contour at $9.43$ km, corresponding to the lower bound from NICER measurements, provides a powerful complementary constraint, ruling out regions of the parameter space where the star would become unrealistically compact.

\begin{figure*}[t!]
  \centering
  \begin{minipage}[b]{0.47\linewidth}
    \centering
    \includegraphics[width=\linewidth]{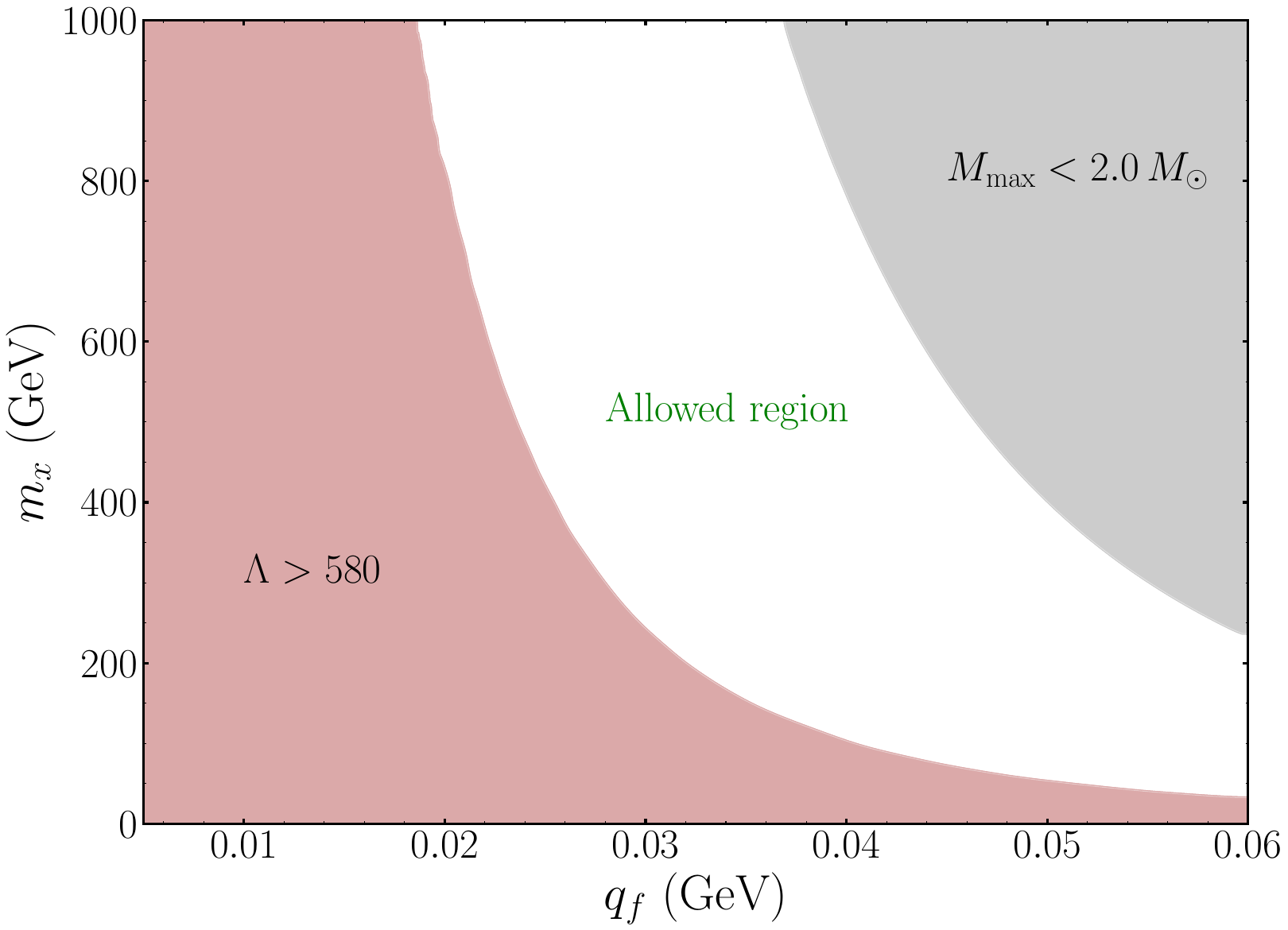}
  \end{minipage}\hspace{0.02\linewidth}%
  \begin{minipage}[b]{0.47\linewidth}
    \centering
    \includegraphics[width=\linewidth]{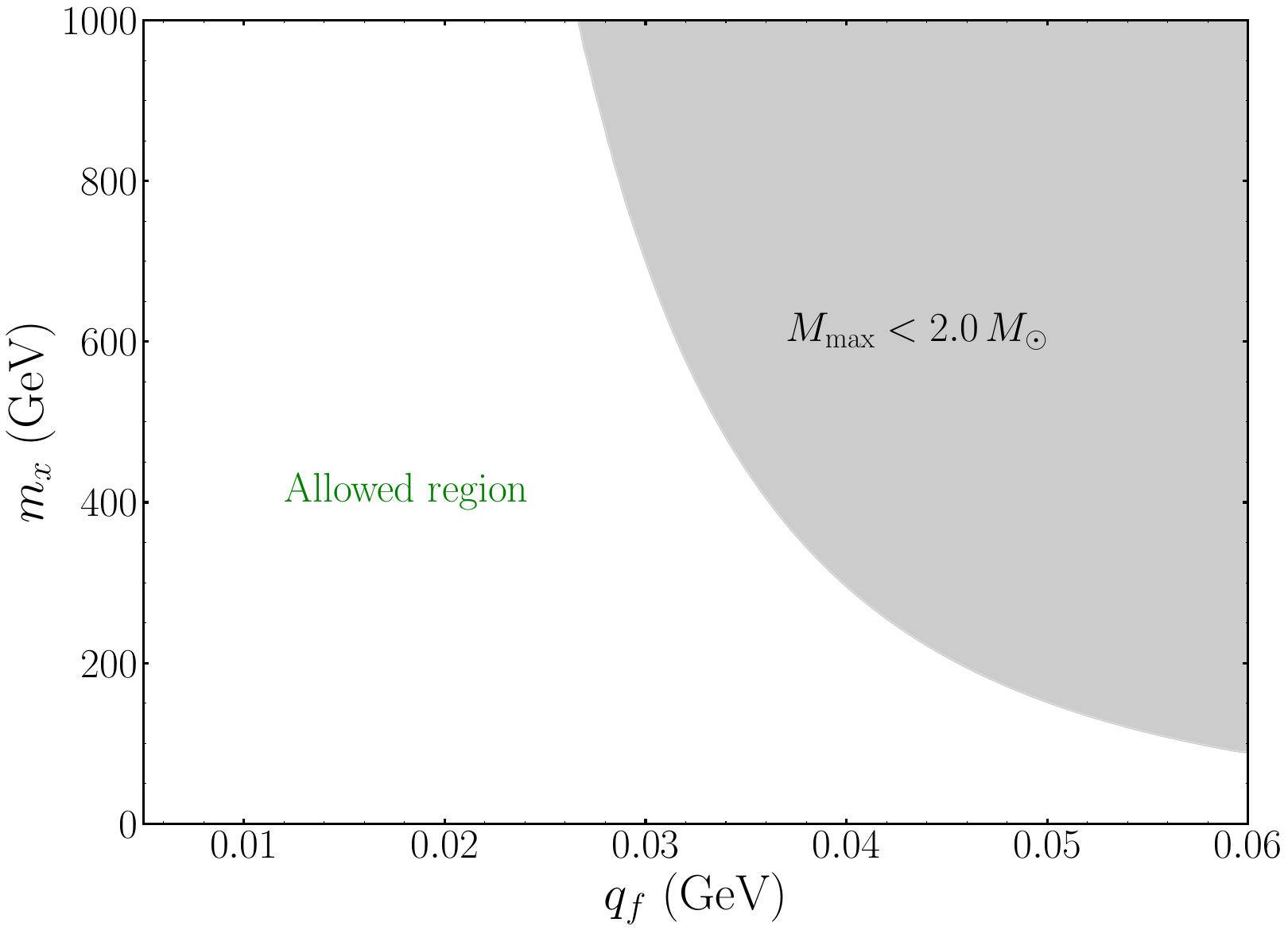}
  \end{minipage}
  \caption{Allowed ALP ($q_f, m_\chi$) parameter space for dark matter admixed stars using the stiff hadronic EoS1 (left) and the soft EoS18 (right). The gray-shaded region denotes the parameter space where the maximum mass falls below the $2\,M_\odot$ observational limit \cite{Demorest:2010bx, Antoniadis:2013pzd, Fonseca:2021wxt, NANOGrav:2019jur}, while the red-shaded region indicates where the dimensionless tidal deformability exceeds the upper bound of 580 from the GW170817 binary neutron star merger \cite{LIGOScientific:2017vwq, LIGOScientific:2018cki}. The white region in each plot represents the allowed parameter space with both constraints.
}
  \label{mr}
\end{figure*}

When compared to the results from the stiff hadronic EoS (Figure~\ref{contour_stiff}), the constraints on the DM parameter space become considerably more stringent. The stiff EoS begins with a much higher maximum mass ($\sim 2.7\,M_\odot$) and a larger radius ($\sim 13.78$ km), providing a substantial buffer against the softening effects of dark matter. For the stiff EoS, only a very large admixture of DM can violate the $2\,M_\odot$ mass limit, leaving most of the parameter space viable. In strong contrast, the soft EoS18, with its baseline $M_{max}$ of only $2.2\,M_\odot$, is far more sensitive. Even a modest DM component can push the maximum mass below the observational threshold, thus ruling out a much larger area of the $q_f - m_\chi$ plane. This demonstrates that astrophysical constraints on dark matter are critically modulated by the underlying properties of the nuclear EoS; softer EoS models, being inherently closer to the observational limits, allow for significantly tighter constraints on the properties of DM.

Figure~\ref{mr} shows the joint constraints on the DM mass $m_\chi$ and the Fermi momentum $q_f$, from the maximum mass and tidal deformability bounds. The left and right plots correspond to the stiff EoS1 and the soft EoS18, respectively. 
The unshaded ``Allowed region'' represents the combinations of DM parameters that are consistent with current astrophysical observations. Two primary observational bounds constrain this region: The first, labeled $M_{max} < 2.0\,M_\odot$ (gray shaded region), excludes the parameter space where the DM admixture softens the EoS to such an extent that the star can no longer support a maximum mass of at least two solar masses, the observational requirement from PSR~J0348+0432 and PSR~J0740+6620 \cite{Demorest:2010bx, Antoniadis:2013pzd}. The second constraint, labeled $\Lambda > 580$ (red shaded region), rules out parameter combinations that would result in a tidal deformability for a $1.4\,M_\odot$ star larger than the upper limit derived from gravitational wave event GW170817 \cite{LIGOScientific:2017vwq, LIGOScientific:2018cki}.

For the stiff EoS (left plot), two distinct exclusion mechanisms shape a well-defined allowed ``island.'' At low $q_f$ (i.e., negligible DM admixture), the models are too extended and yield $\Lambda_{1.4} > 580$ (red region). This red boundary recedes as $q_f$ increases and as $m_\chi$ becomes larger, reflecting the tendency of dark matter to reduce stellar radii and tidal deformabilities. At the same time, increasing $q_f$ lowers the maximum mass, which carves out the gray, upper-right exclusion region associated with $M_{\rm max} < 2\,M_\odot$. Between these two exclusion zones lies a well-defined allowed region, constrained to $0.02 \leq q_f \leq 0.06$ and $43\,\mathrm{GeV} \leq m_\chi \leq 1000\,\mathrm{GeV}$. So for a given stiff EoS, these constraints only exclude a relatively small portion of the parameter space at high DM coupling values. Note that while this figure defines the strictly allowed physical region, the relative likelihood of parameters within the white region are not equally probable. That can be seen in Figure \ref{fig:posterior_mx_qf} for the marginalized probability distributions.

For the soft EoS (right plot), the tidal deformability constraint does not become active (no red region) because these models are already compact. Instead, the $2\,M_\odot$ requirement dominates: even modest DM admixtures further suppress the maximum mass, so most of the plane is excluded by $M_{\rm max} < 2\,M_\odot$ (gray region). The allowed region is therefore a narrow sliver confined to small $q_f$ and relatively heavier $m_\chi$, illustrating that soft EoSs are less compatible with significant dark matter content. Overall, the figure highlights that the viable ALP-mediated DM parameter space is highly sensitive to the underlying nuclear EoS: for stiff EoSs, both $\Lambda_{1.4}$ and $M_{\rm max}$ act in concert to localize an allowed band, whereas for soft EoSs the $2\,M_\odot$ bound alone leaves only a very restricted region at low $q_f$. This demonstrates the discriminating power of combining mass and tidal-deformability constraints when probing DM physics in NSs.

A direct comparison of the two figures reveals a crucial observation: the allowed parameter space for dark matter is dramatically larger when a soft hadronic EoS is considered. The stiff EoS is inherently more robust against the gravitational burden of a DM core, and as a result, a small region of the $q_f - m_\chi$ plane remains consistent with observations. In contrast, the soft EoS is far more sensitive to the presence of DM. Its baseline properties are already closer to the observational limits, and therefore, even a modest DM fraction is sufficient to push the star's maximum mass or radius into the excluded zone. This difference underscores that astrophysical constraints on dark matter properties are not absolute but are critically dependent on the underlying physics of dense nuclear matter. 

\begin{figure*}[htbp]
  \centering
  \begin{minipage}[b]{0.47\linewidth}
    \centering
    \includegraphics[width=\linewidth]{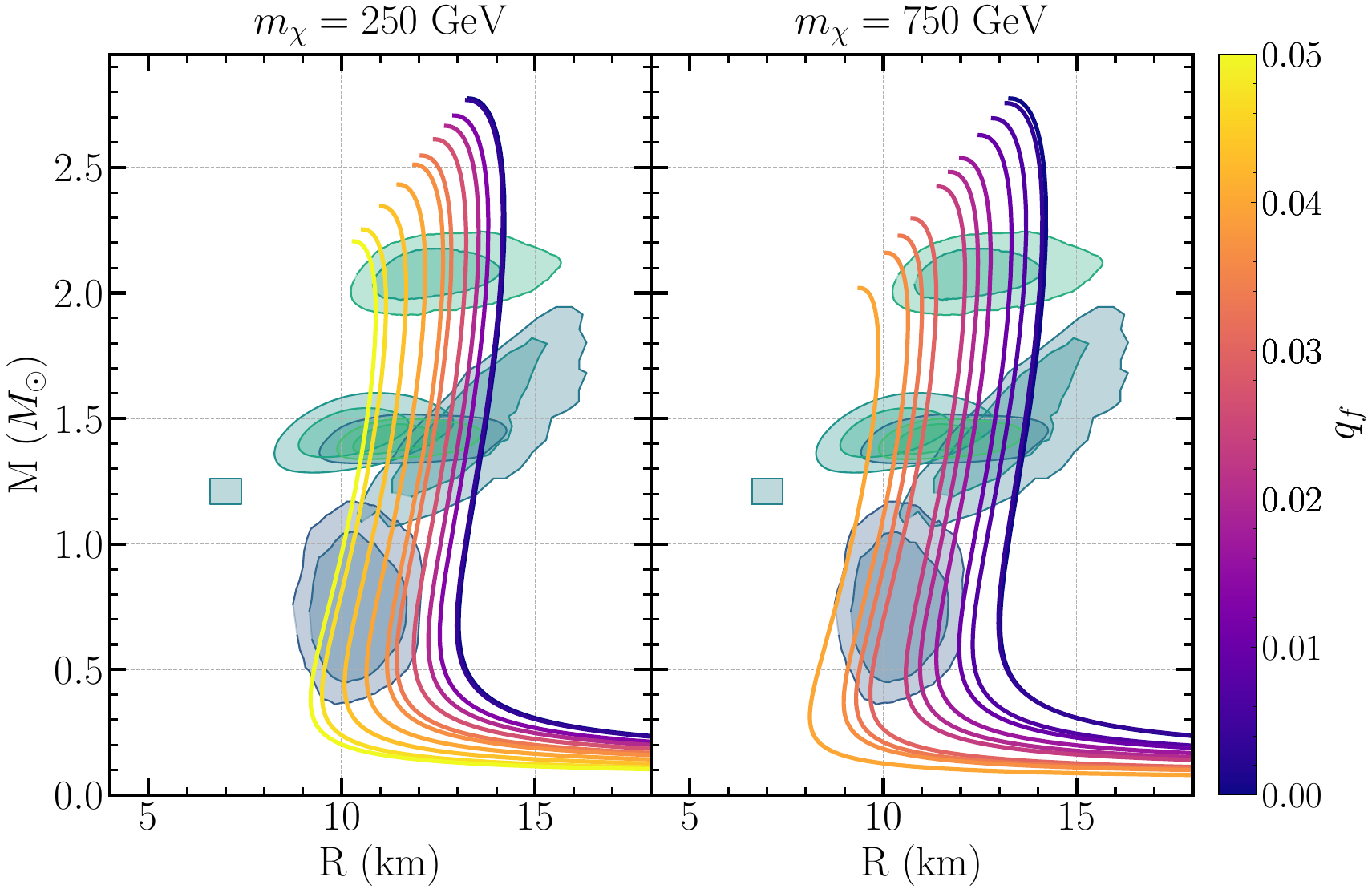}
  \end{minipage}\hspace{0.02\linewidth}%
  \begin{minipage}[b]{0.47\linewidth}
    \centering
    \includegraphics[width=\linewidth]{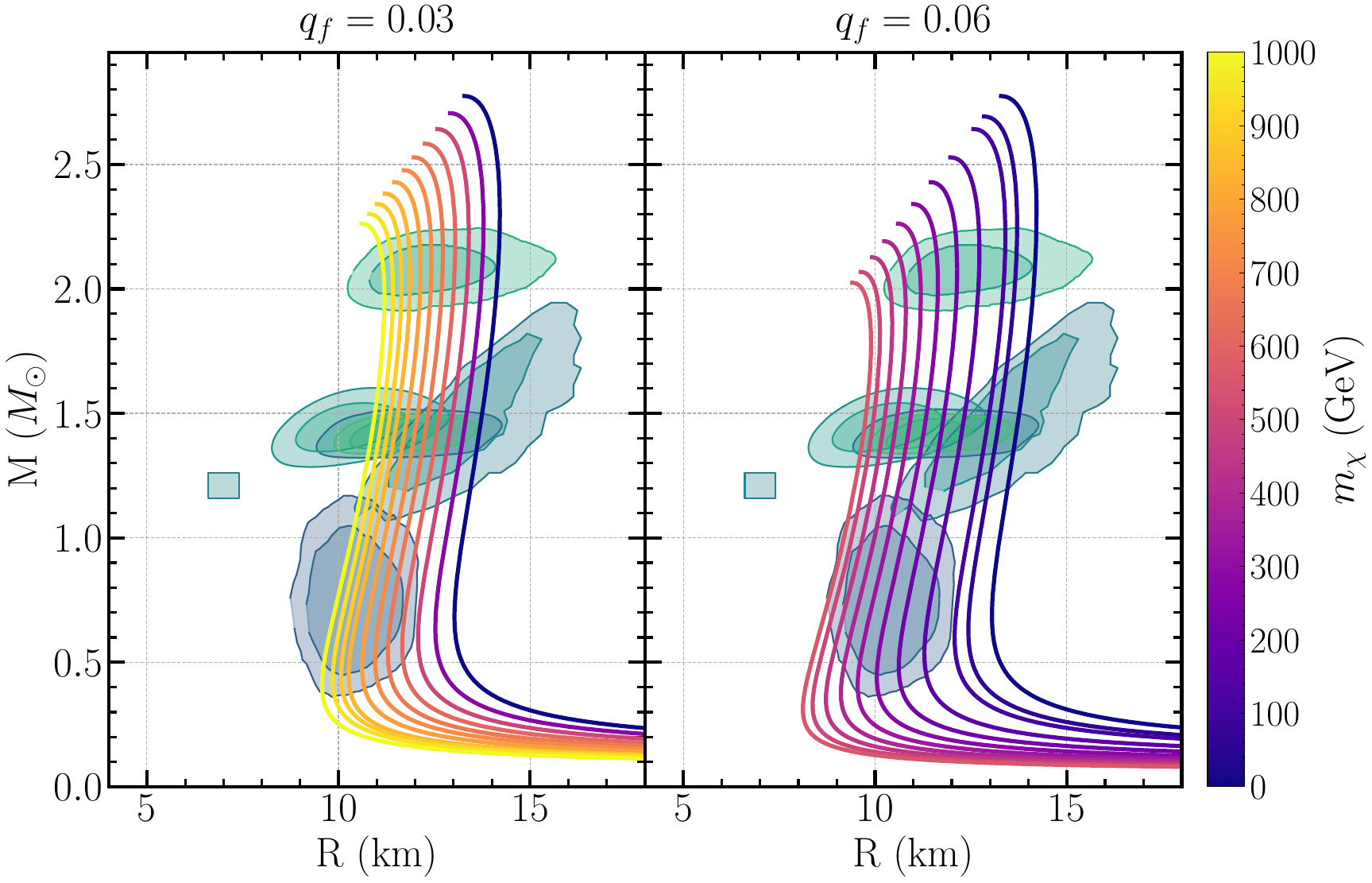}
  \end{minipage}
  \caption{Selected $m_\chi$ and $q_f$ parameter combinations with stiff hadronic EoS (EoS1) that satisfy all the astrophysical constraints. Left plot: MR relations at fixed $m_\chi$ = 250 and 750 GeV, respectively, with colorbar representing the varying $q_f$ parameter. Right plot: MR relations at fixed $q_f$ = 0.03 and 0.06 GeV, respectively, with colorbar representing the varying $m_\chi$ parameter. Shaded regions represent the observational constraints, same as in Figure \ref{MR_fix}.}
  \label{MR_selected}
\end{figure*}

A stiffer EoS allows for significantly more stringent limits to be placed on the nature of dark matter. For this reason, we focus on dark matter models employing a stiff hadronic EoS, as they allow us to better constrain the dark matter parameter space within the range permitted by astrophysical observations.

With the allowed region of DM parameters established, we now examine the corresponding mass–radius relations. In particular, we plot the selected MR curves at varying DM parameter values to assess their consistency with astrophysical constraints.

Figure~\ref{MR_selected} shows the MR relations for DM-admixed NS model with stiff hadronic EoS18, that satisfy all astrophysical constraints described in Section~\ref{astro_const}. The left plot with two panels corresponds to fixed DM particle masses (250~GeV and 750~GeV) with varying Fermi momentum $q_f$, while the right plot represents two panels with fixed $q_f = 0.03$ and $0.06$~GeV with varying $m_\chi$. The allowed MR relations are significantly narrowed compared to the full set, highlighting only those parameter combinations that lie within the observational posteriors of key pulsars, including PSR~J0348+0432, PSR~J0030+0451, PSR~J0740+6620, PSR~J0437--4715, and PSR~J0614--3329. One important feature is that any model satisfying the PSR~J0614--3329 constraint automatically satisfies the HESS~J1731--347 constraint, indicating a nested compatibility between these objects within the filtered parameter space and that the HESS J1731--347 could be an exotic object as shown in the literature \cite{Rather:2023tly, Pitz:2024xvh, Gholami:2024ety, Sagun:2023rzp, Oikonomou:2023otn}. Another noteworthy result is that, under the stiff EoS, the strict constraint filtering imposes a lower bound on the dark matter particle mass, such that $m_\chi$ must be greater than 43~GeV. 

Figure~\ref{MR_fix} indicated that several MR relations satisfy the observational region of XTE~J1814--338, particularly at higher values of both $m_\chi$ and $q_f$. However, these configurations fail to satisfy the $2\,M_\odot$ constraint, which is a fundamental requirement for any viable EoS. After imposing the observational constraints and restricting the DM parameter space accordingly, none of the resulting MR relations are able to reproduce the XTE~J1814--338 measurement, as they remain incompatible with the two-solar-mass limit. In contrast, the same set of models successfully accommodates the low-mass compact object HESS J1731--347. This result highlights that the proposed DM framework is capable of simultaneously accounting for massive NSs with $M \sim 2\,M_\odot$ and for low-mass compact objects within a unified and self-consistent description.

A key challenge in constraining exotic matter is to disentangle its signature from the inherent uncertainties in the baryonic EoS. To quantitatively assess whether the DM effect is distinguishable from the uncertainty of the nuclear EoS, we performed a direct comparison. We define the \emph{signal-to-uncertainty ratios} as
$S_R=\big|\Delta R_{1.4}^{\mathrm{DM}}\big|/{\sigma_R}$ and
$S_\Lambda=\big|\Delta\Lambda_{1.4}^{\mathrm{DM}}\big|/{\sigma_\Lambda}$,
where $\Delta R_{1.4}^{\mathrm{DM}}=R_{1.4}^{\mathrm{DM}}-R_{1.4}^{0}$ and
$\Delta\Lambda_{1.4}^{\mathrm{DM}}=\Lambda_{1.4}^{\mathrm{DM}}-\Lambda_{1.4}^{0}$. $\sigma_R=0.54~\text{km}$ and 
$\sigma_\Lambda=178.5$ is defined as the \emph{across-model} nuclear uncertainty as half the difference between our two purely nucleonic EoSs (see Table \ref{tab:model_metrics}).
 We find that the DM effects remain subleading ($S<1$) for small fractions (\(q_f \lesssim 0.02\)) but dominate ($S>1$) the uncertainty budget for moderate admixtures (\(q_f \gtrsim 0.03\)). This provides a clear and quantitative way to assess when DM effects become distinguishable from the intrinsic nuclear-EoS uncertainty.


\subsection{Statistical analysis}
\begin{figure*}[htbp]
  \centering

  \begin{minipage}[b]{0.47\linewidth}
    \centering
    \includegraphics[width=\linewidth]{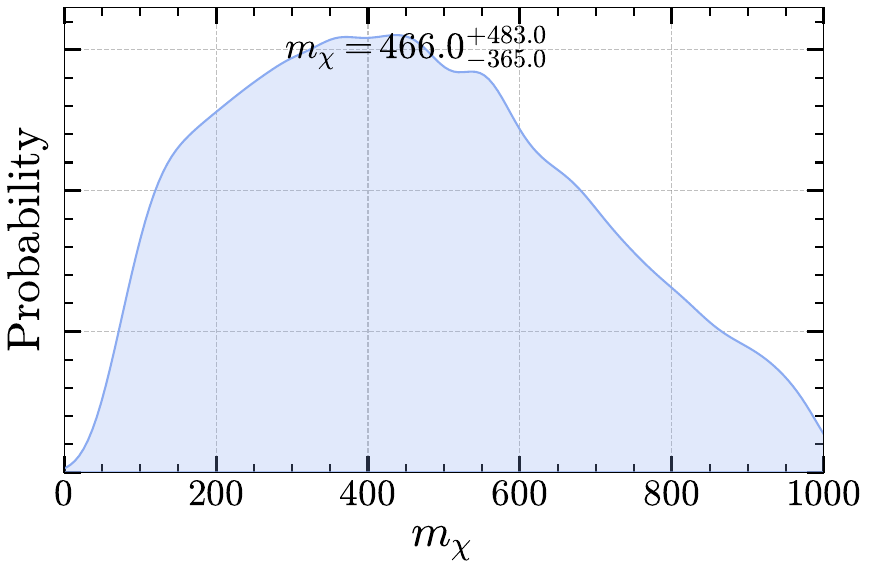}
    
  \end{minipage}\hspace{0.04\linewidth}%
  \begin{minipage}[b]{0.47\linewidth}
    \centering
    \includegraphics[width=\linewidth]{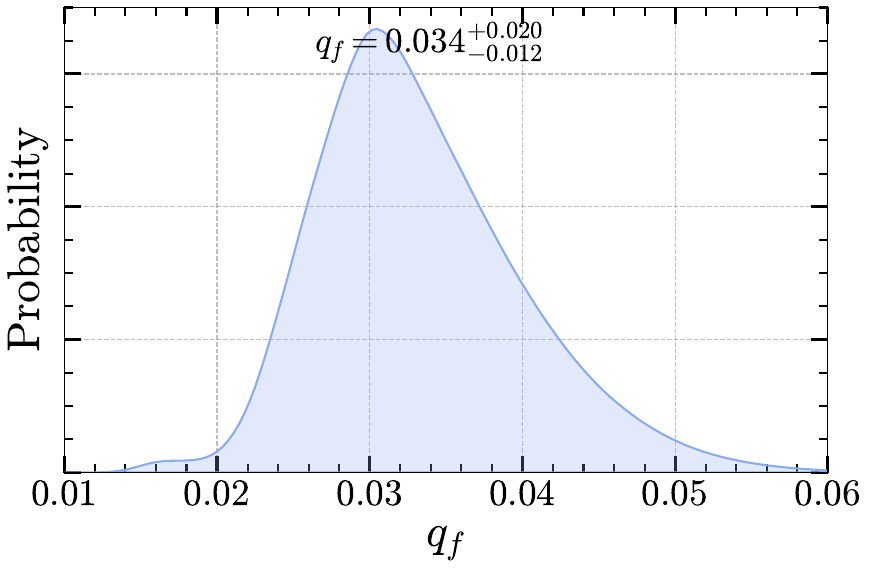}
    
  \end{minipage}

  \caption{The probability distributions for the dark matter particle mass $m_\chi$ (left) and the dark matter fermi momentum $q_f$ (right) for EoS1, inferred from mass--radius constraints. 
The median values with 90\% credible intervals are $m_\chi = 466^{+483}_{-365}$ and $q_f = 0.034^{+0.020}_{-0.012}\,\mathrm{GeV}$.}
  \label{fig:posterior_mx_qf}
\end{figure*}

After applying the astrophysical constraints described in the previous section, we proceed to the statistical analysis. As outlined earlier, each EoS group (EoS1 and EoS 18) contains 30,601 parameter sets, which were individually tested against the astrophysical criteria. Among these, 16,524 models from the soft EoS group satisfied all the constraints, whereas only 5,746 models passed in the case of the stiff EoS. Each surviving model was then evaluated and assigned a composite score based on the following three components:

Each passing model was then evaluated and assigned a composite score based on the following three components:
\begin{itemize}
  \item A \textbf{voting score}, reflecting how many constraints the model satisfies within $1\sigma$ (full weight) or $2\sigma$ (half weight);
  \item A \textbf{likelihood-based score}, based on the Gaussian-weighted proximity of the model's MR relation to the centers of the constraint regions;
  \item A \textbf{KDE-based score}, which quantifies the agreement between the model's MR relation and the overall observational distribution using kernel density estimation.
\end{itemize}

All three scores were normalized and combined using fixed weights (30\% voting, 40\% likelihood, and 30\% KDE) to yield a final ranking for each model. The filtered dataset, consisting of observationally consistent EoS models and their full mass-radius curves, was saved for statistical analyses. The above scheme was chosen to balance strict constraint satisfaction with quantitative goodness-of-fit (likelihood, KDE). This ensures that both binary agreement and statistical proximity are reflected in the scoring. Reasonable variations of these weights yield consistent trends, confirming the robustness of the filtering scheme.

Figure~\ref{fig:posterior_mx_qf} presents the score-weighted probability distributions for the DM particle mass $m_\chi$ (left panel) and the DM Fermi momentum  
$q_f$ (right panel), obtained from the subset of NS models constructed using EoS1 that satisfy all astrophysical constraints under the filtering scheme, including HESS and PSR J0614-3329. These histograms represent the marginalized posterior probabilities for each parameter individually, meaning the distributions correspond to the projection of the valid two--dimensional parameter space onto a single axis. Consequently, a high probability density for a specific value does not imply that all corresponding pairings in the $(m_\chi, q_f)$ plane are physically valid; the allowed joint parameter space is strictly constrained by the exclusion regions shown in Figure~\ref{mr}. The left panel shows that the distribution for $m_\chi$ spans a broad range, with a relatively flat profile between approximately 200 and 700\,GeV, and a peak centered at $m_\chi = 466\,\mathrm{GeV}$. The 90\% credible interval extends from 101\,GeV to 949\,GeV ($^{+483}_{-365}\,\mathrm{GeV}$), indicating that while intermediate-mass DM particles are mildly preferred, the data remain consistent with a wide range of values. The shape of the distribution, with a slow fall-off at higher masses, reflects the intrinsic degeneracy in the particle mass when constrained by NS observables.

The right panel displays a much more sharply peaked distribution for $q_f$, with a best-fit value of $q_f = 0.034$ and a 90\% credible interval of $^{+0.020}_{-0.012}$. The distribution rises steeply near $q_f \sim 0.02$, peaks at approximately $q_f \sim 0.034$, and declines gradually toward the upper bound of the scanned range. This asymmetry suggests a preference for a small but non-zero DM-admixture in NSs, while disfavoring both very low and high values. The narrow width of the distribution indicates that $q_f$ is more tightly constrained than $m_\chi$, especially in the context of stiff EoS, which are capable of supporting high-mass NS configurations. These results identify the region of the DM parameter space that are most consistent with the NS observations when evaluated using stiff EoS. The combination of constraint filtering and physically motivated EoS models leads to a clear preference for moderate-mass DM particles and low DM fractions that still impact the macroscopic structure of NSs. In Table~\ref{tab:EoS_median_ci_block}, we report the medians and 90\% confidence intervals (CIs) of key observables for the soft and stiff EoS models. 
For the soft EoS, the maximum mass is $2.24\,M_\odot$ with a radius of $11.19$ km, while the stiff EoS yields a higher $M_{\max}=2.43\,M_\odot$ and $R_{\max}=11.47$ km. Radii at benchmark masses ($0.77\,M_\odot$, $1.4\,M_\odot$, and $2.08\,M_\odot$) and the tidal deformability at $1.4\,M_\odot$ show the same trend: the stiff EoS consistently predicts larger values, with $\Lambda_{1.4}=466$ compared to $379$ for the soft case.

\begin{table}[htbp]
\begin{minipage}{\linewidth}
\centering
\caption{Median and 90\% confidence intervals (CI) for selected observables from the soft and stiff EoS models. Masses are in units of $M_\odot$, radii are in km, and $\Lambda$ is the dimensionless tidal deformability.}
\vspace{0.5em}
\begin{tabular}{l p{1.0cm}c p{1.0cm}c}
\toprule
Quantity
& \multicolumn{2}{c}{EoS\_soft} 
& \multicolumn{2}{c}{EoS\_stiff} \\
\cmidrule(lr){2-3} \cmidrule(lr){4-5}
& Med. & 90\% CI & Med. & 90\% CI \\
\midrule
$M_\mathrm{max}$    
& 2.24 & [2.11, 2.26]    
& 2.43 & [2.28, 2.70] \\

$R_\mathrm{max}$    
& 11.19 & [10.46, 11.31] 
& 11.47 & [10.69, 12.11] \\

$R_{0.77}$          
& 10.90 & [10.10, 11.05]  
& 10.85 & [10.06, 11.58] \\

$R_{1.4}$           
& 11.84 & [11.02, 11.97] 
& 11.70 & [10.93, 12.34] \\

$R_{2.08}$          
& 11.94 & [9.89, 12.10]  
& 12.16 & [11.27, 12.85] \\

$\Lambda_{1.4}$     
& 379 & [260, 470]
& 466 & [300, 800] \\

\bottomrule
\end{tabular}
\label{tab:EoS_median_ci_block}
\end{minipage}
\end{table}

\begin{figure*}[htbp]
  \centering
  \begin{minipage}[b]{0.50\linewidth}
    \centering
    \includegraphics[width=\linewidth]{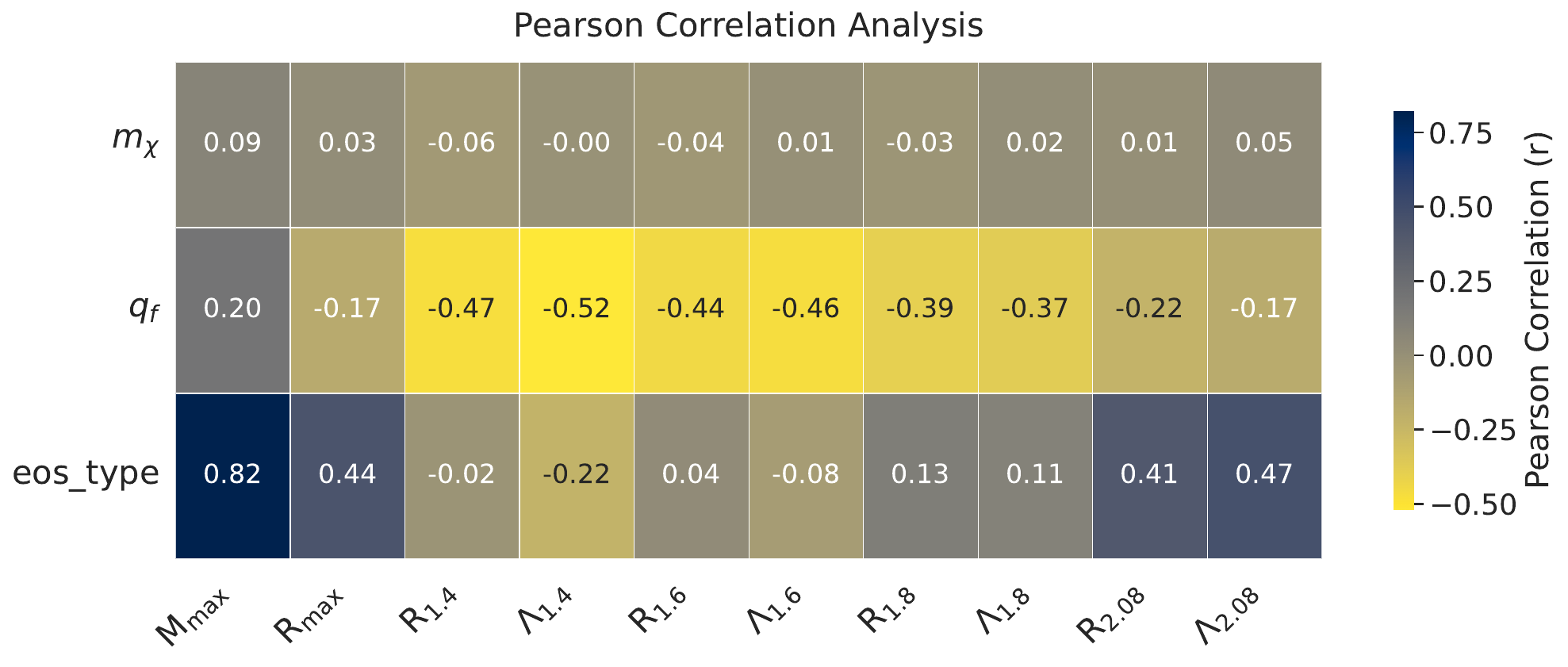}
  \end{minipage}\hspace{0.02\linewidth}%
  \begin{minipage}[b]{0.47\linewidth}
    \centering
    \includegraphics[width=\linewidth]{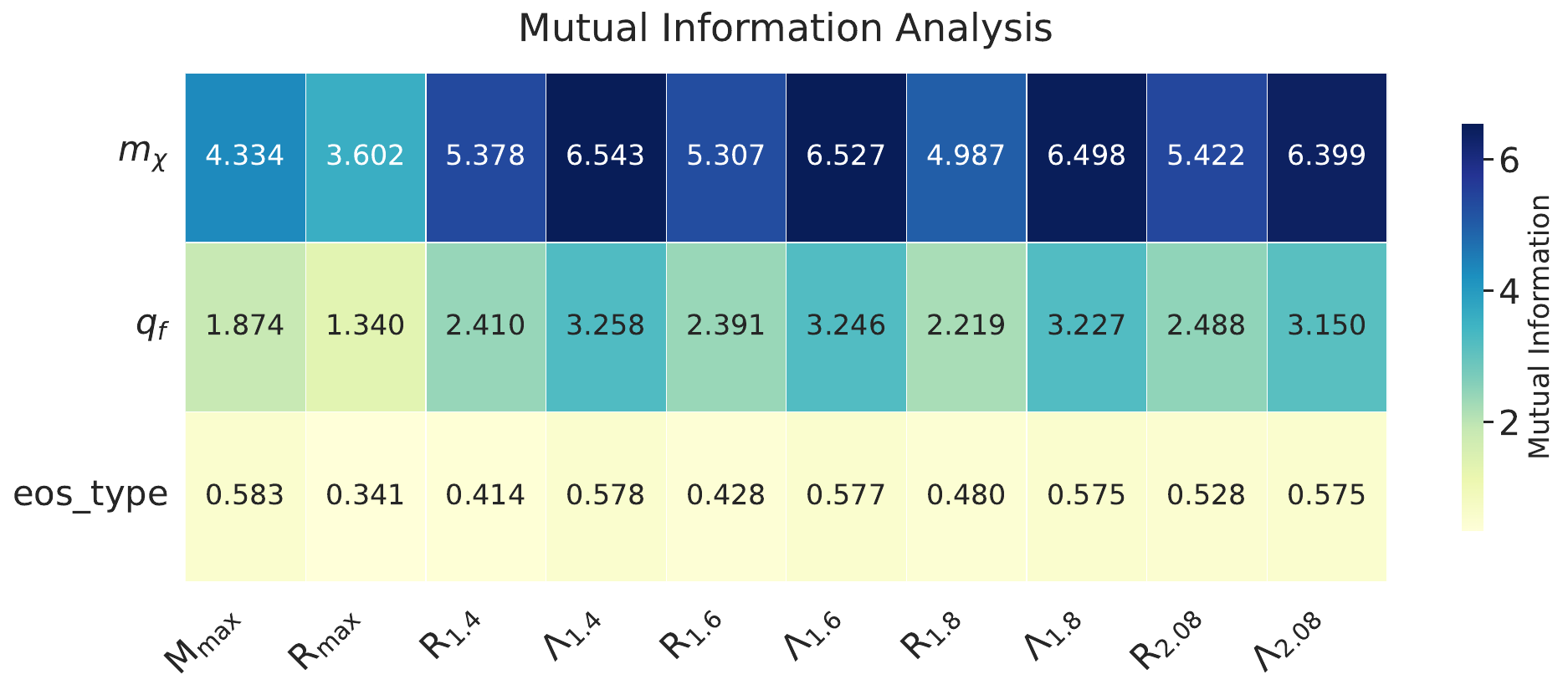}
  \end{minipage}
  \caption{
  Correlation analysis between model parameters and observable quantities. 
  Left panel: Pearson correlation coefficients indicating linear relationships. 
  Right panel: Mutual information scores capturing both linear and nonlinear dependencies.
  }
  \label{fig:correlation_soft}
\end{figure*}

\begin{figure*}[htbp]
  \centering
  \begin{minipage}[b]{0.70\linewidth}
    \centering
    \includegraphics[width=\linewidth]{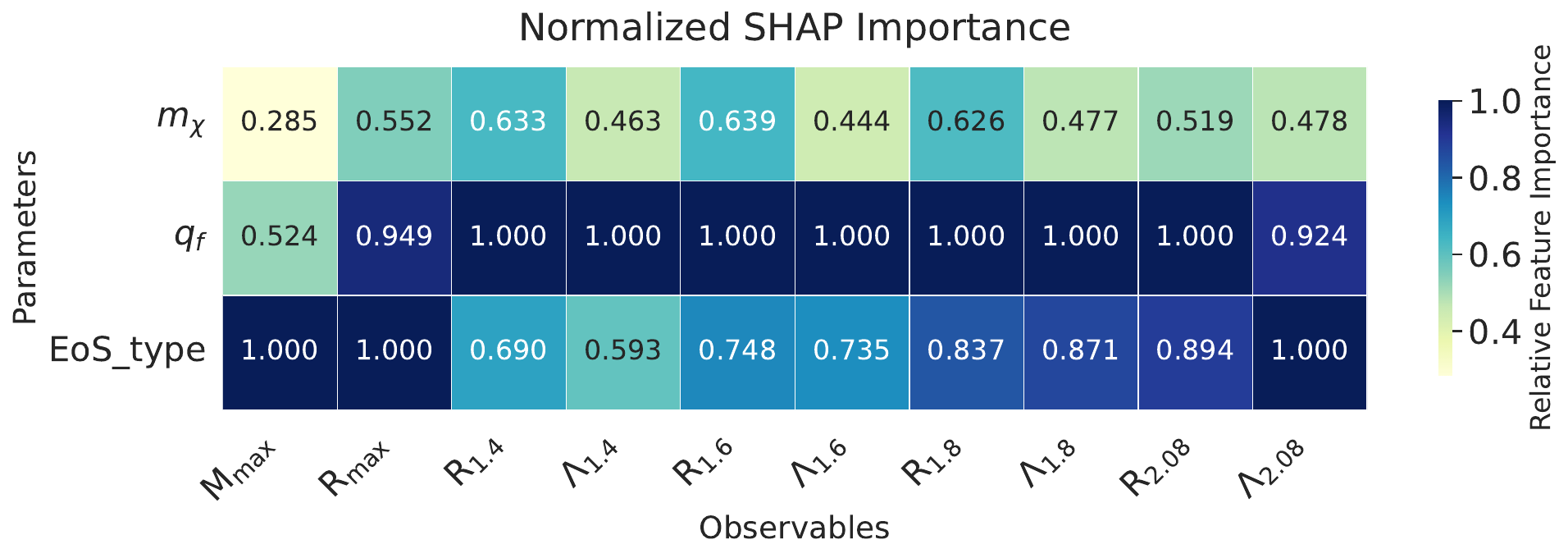}
  \end{minipage}\hspace{0.02\linewidth}%

  \caption{Heatmap of the normalized SHAP feature importance from the XGBoost model, showing the relative influence of the model parameters on the neutron star observables.}
  \label{fig:shap_corelation}
\end{figure*}

Our investigation into the parameter dependencies began with a Pearson correlation analysis on the combined soft and stiff EoS dataset, which explicitly included the \texttt{eos\_type} as a feature (Figure~\ref{fig:correlation_soft}). Here, \texttt{eos\_type} is a binary indicator variable (0 = soft, 1 = stiff). A positive Pearson coefficient implies that the stiff EoSs tend to yield larger values of the corresponding observable, whereas a negative coefficient indicates the opposite. The results reveal that the \texttt{eos\_type} exhibits the strongest association with the maximum mass, with a Pearson correlation of $r = 0.82$, indicating that stiffer EoSs are associated with higher $M_{\mathrm{max}}$ on average. For the dark matter parameters, DM fermi momentum ($q_f$), shows a moderate negative linear correlation with several observables, particularly the tidal deformabilities ($r \approx -0.52$). In contrast, the DM mass ($m_\chi$), exhibits no significant linear correlation with any NS observable. This observation aligns with the findings of \citet{Thakur:2023aqm}, where they investigated two-fluid models of fermionic DM-admixed NSs and concluded that incorporating uncertainties in the baryonic sector diminishes the strength of linear correlations between the DM parameters and their stellar properties. These results suggest that while the \texttt{eos\_type} and $q_f$ are the primary drivers of any observable linear trends, the overall weakness of such correlations emphasizes the need for more sophisticated, non-linear techniques to fully capture the intricate dependencies among model parameters and astrophysical observables.

To probe for all types of statistical dependence, including non‐linear relationships, we next performed a Mutual Information (MI) analysis. Mutual Information is a powerful metric from information theory that quantifies the “amount of information” obtained about one random variable through observing another. It measures the reduction in uncertainty of an observable (e.g., $M_{\max}$) given the knowledge of a parameter (e.g., $m_\chi$). A higher MI value signifies a stronger statistical relationship of any kind, not just linear. The results of this analysis, presented in the right panel of Figure~\ref{fig:correlation_soft}, reveal a clear hierarchy of statistical dependence. The DM particle mass, $m_\chi$, exhibits by far the strongest dependence, with MI scores consistently above 3.6 and reaching as high as 6.5. This indicates that $m_\chi$ shares the most information with the observables and that its relationship with them is both strong and complex. Following this, $q_f$ shows a moderate dependence, with MI scores generally ranging from 1.3 to 3.2. For context, an MI value of 6.5 conveys approximately twice the shared information compared to 3.2, underscoring a substantially stronger statistical dependence. This confirms a significant, though secondary, statistical relationship. Most strikingly, the \texttt{eos\_type} feature registers very low MI scores (all below 0.6). This result appears to directly contradict the strong linear correlation we observed with $M_{\max}$ in Pearson correlation. This discrepancy is not an error but rather a known characteristic of the MI metric itself: MI is sensitive to the entropy, or inherent variability, of a feature. A continuous variable like $m_\chi$, which can take on many values, has high entropy and thus a large capacity to reduce the uncertainty of an observable. In contrast, a simple binary feature like \texttt{eos\_type} (only 0 or 1) has very low entropy, limiting its mathematical ability to reduce the uncertainty of a continuous observable and resulting in a lower MI score regardless of its true physical importance. This ambiguity underscores the limitation of relying on a single statistical measure and necessitates the use of a model‐based approach like SHAP to determine the practical predictive importance of each feature. \\
To resolve the ambiguities from purely statistical methods and to quantify the predictive importance of each parameter, we employed a robust machine learning approach. For each NS observable, we trained an XGBoost model \cite{Chen_2016}, a powerful gradient‑boosted decision tree algorithm, using the DM parameters ($m_\chi$, $q_f$) and the \texttt{EoS\_type} as input features. To ensure the model’s reliability, we evaluated its performance on a held‑out test set, achieving high R‑squared (\textbf{R$^2$}) scores for all observables (e.g., 0.9998 for $M_{\max}$ and 0.8030 for $\Lambda_{1.4}$), confirming that the models learned the underlying relationships with high fidelity. We then used the SHAP (SHapley Additive exPlanations)~\cite{lundberg2017unifiedapproachinterpretingmodel} framework to explain the predictions of these accurate models. The resulting feature importances were normalized for each observable, assigning a score of 1.0 to the most influential parameter. The results, visualized in Figure~\ref{fig:shap_corelation}, reveal a clear and physically intuitive picture. The EoS type is the dominant factor for determining global structural properties, such as the maximum mass ($M_{\max}$) and the radius of high‑mass stars ($R_{\max}$, $R_{1.8}$, $R_{2.08}$). In contrast, $q_f$ emerges as the primary driver for the tidal deformability ($\Lambda$) across all mass ranges and for the radius of a canonical $1.4\,M_\odot$ star. The DM particle mass ($m_\chi$) consistently plays a significant but secondary role, modulating the effects of the other two features. This comprehensive analysis definitively shows that the impact of dark matter is not an isolated effect but a complex interplay between the dark matter properties and the underlying nuclear equation of state, with different parameters dominating different physical observables.


\section{Inferring ALP DM Parameters from supervised regression}\label{machinelearning}

The central motivation for this part of the work is to explore machine learning as a rapid and accurate approach to complement traditional methods in inferring ALP-mediated dark-matter properties. Given a set of neutron-star observables derived from the mass--radius curve (our features) and the corresponding dark-matter parameters (our labels), the task reduces to learning the inverse mapping between them. This constitutes a supervised interpolation problem within a high-dimensional parameter space.

Our aim, therefore, is to train a model that interprets the geometry of the MR curve derived from the growing population of NS observations. To achieve this, we employed a random row-wise training strategy. Our extensive simulations create a dense grid covering the physically relevant parameter space ($m_\chi \in [0, 1000]~\mathrm{GeV}$, $q_f \in [0.00, 0.06]~\mathrm{GeV}$). Since our dataset contains millions of configurations densely sampling this domain, we assume that any future astrophysical observation will statistically correspond to a physical configuration that falls within the manifold spanned by our training set. In this context, learning the local mapping $(M, R, \Lambda) \rightarrow (m_\chi, q_f)$ is the appropriate task.


\begin{figure}[htbp]
  \centering
  \begin{minipage}[b]{0.48\linewidth}
    \centering
    \includegraphics[width=1.5\linewidth]{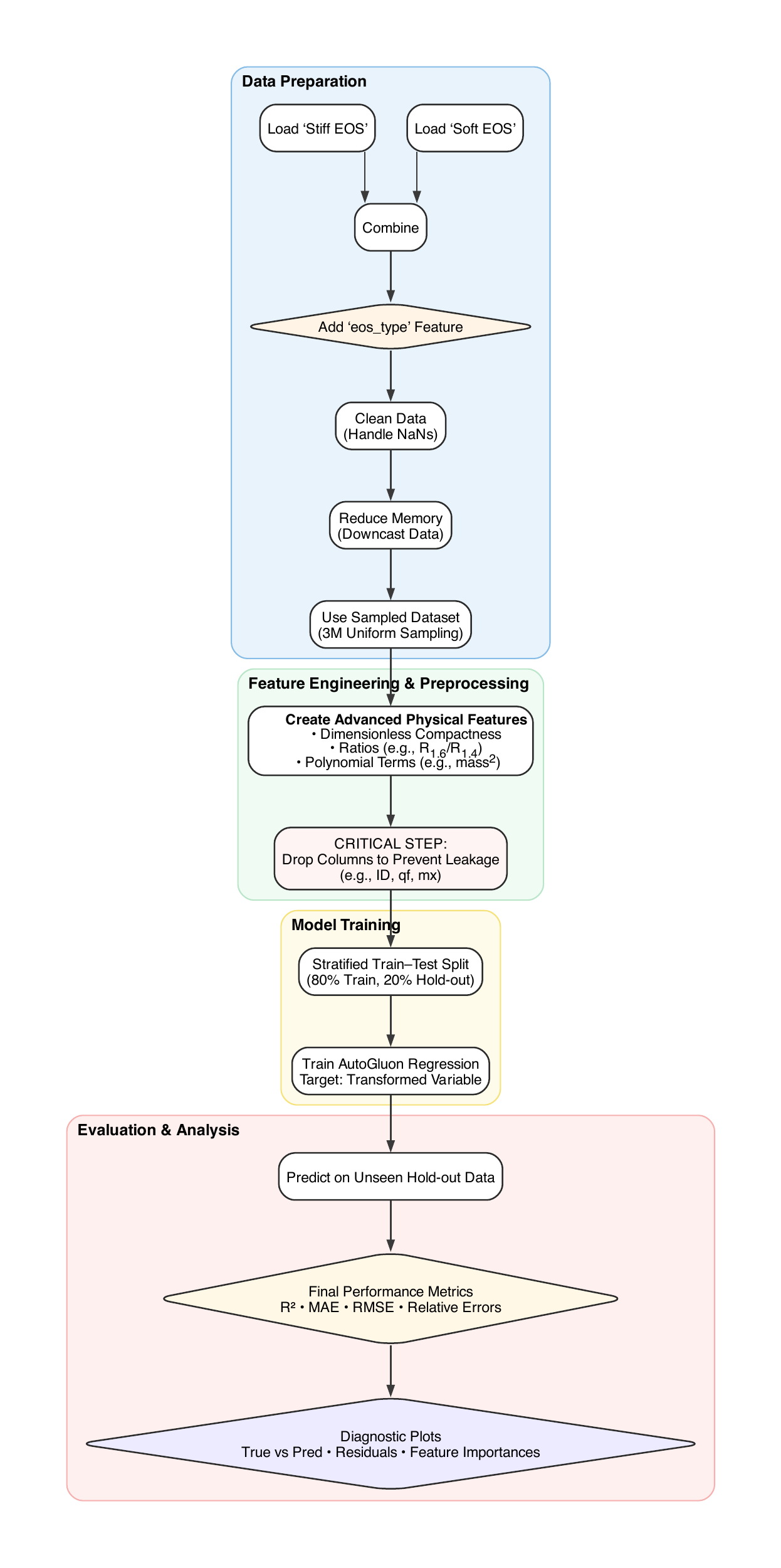}
  \end{minipage}\hspace{0.02\linewidth}%
  \caption{Flowchart depicting the machine learning pipeline for supervised regression used in this work.}
  \label{ML_0}
\end{figure}

 The entire process, from data preparation to final model evaluation, is summarized in the flowchart in Figure~\ref{ML_0}.

Data Preparation and Preprocessing:- To develop a model relating physical observables to DM parameters, we merged datasets corresponding to ``stiff'' and ``soft'' EoS and introduced a categorical \texttt{EoS\_type} feature. This allowed a balanced train-test split and prevented model bias, enhancing the ability of gradient-boosted trees to characterize different data sub-groups. The resulting structure enabled a more optimally tuned weighted ensemble in AutoGluon~\cite{agtabular}. To manage the large dataset (over 11 million rows), we reduced its memory footprint by approximately 46\% through downcasting of numerical data types~\cite{mckinney-proc-scipy-2010} without loss of precision. Subsequently, model training was conducted on a representative uniform sample of 3 million rows to overcome memory limitations while still capturing key non-linear relationships. Standard preprocessing steps included handling of missing (NaN) values and duplicate entries. To avoid data leakage, confounding columns such as dataset identifiers and alternative target variables (e.g., $q_f$ when predicting $m_\chi$) were explicitly dropped prior to the train-test split.

Feature Engineering and Target Transformation:- 
Initial model training with raw observables stagnated, so we introduced engineered features, including compactness values ($C_{1.4}$), stiffness ratios ($R_{1.6}/R_{1.4}$), and lambda ratios ($\Lambda_{1.6}/\Lambda_{1.4}$). These ratios emerged as the most predictive features, as they better encode the EoSs stiffness and its influence on macroscopic observables. Other feature engineering attempts failed to improve, and sometimes even degraded, performance. To address extreme outliers in target variables such as $m_\chi$ (spanning several orders of magnitude) and $q_f$ (clustered near zero), we applied a logarithmic transformation. This crucial step shifted the model's optimization objective from minimizing absolute error to minimizing relative error, thereby suppressing outliers by enabling order-of-magnitude corrections on the original scale. We trained our models using the AutoGluon tabular framework~\cite{agtabular}, an AutoML library that builds a robust, multi-layer weighted ensemble through techniques like bagging~\cite{breiman1996bagging} and stacking~\cite{Wolpert_1992}. The final, processed dataset was split into an 80\% training set and a 20\% hold-out test set for final validation.

\begin{figure*}[t!]
  \centering
  \begin{minipage}[b]{0.47\linewidth}
    \centering
    \includegraphics[width=\linewidth]{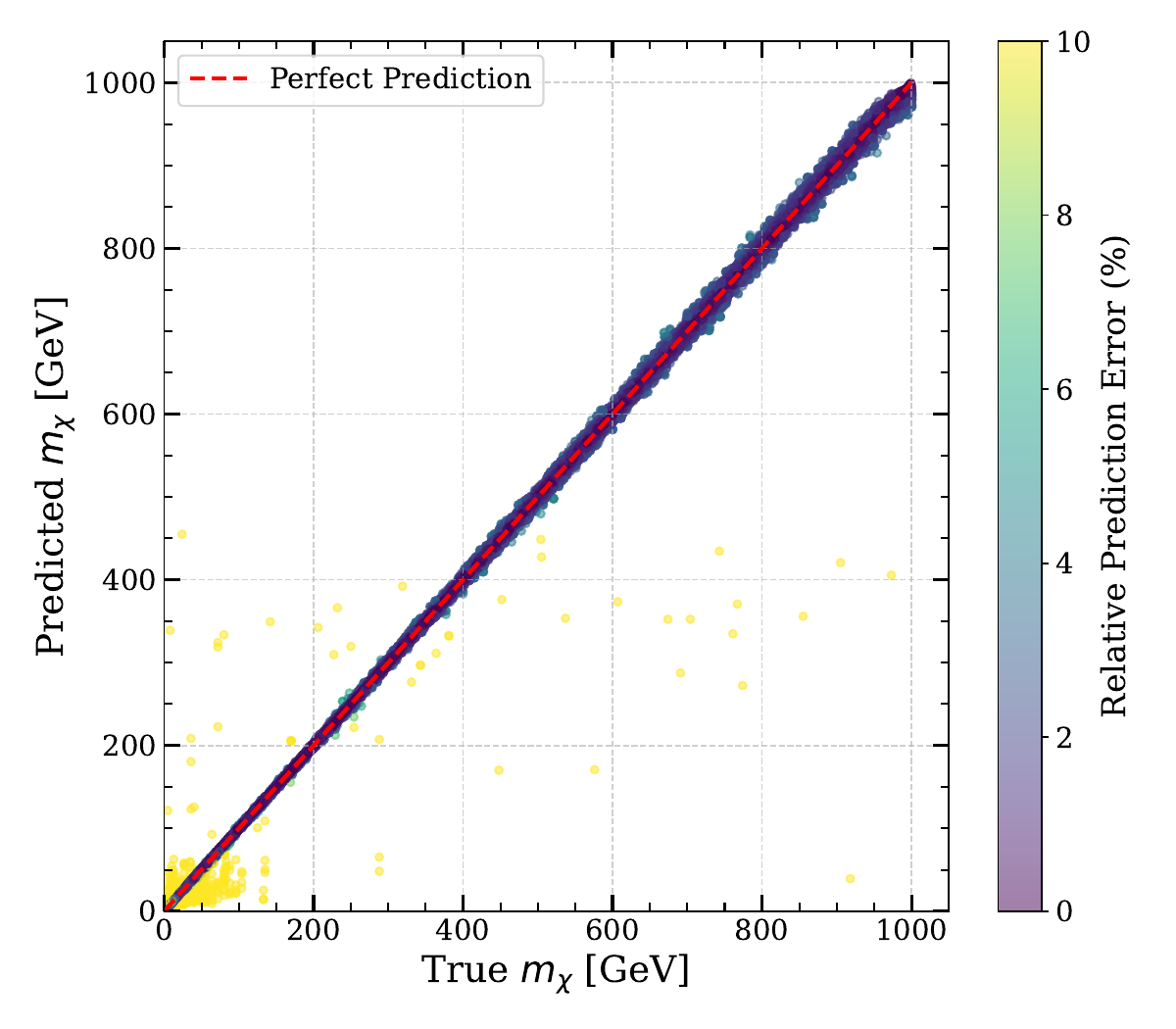}
  \end{minipage}\hspace{0.02\linewidth}%
  \begin{minipage}[b]{0.47\linewidth}
    \centering
    \includegraphics[width=\linewidth]{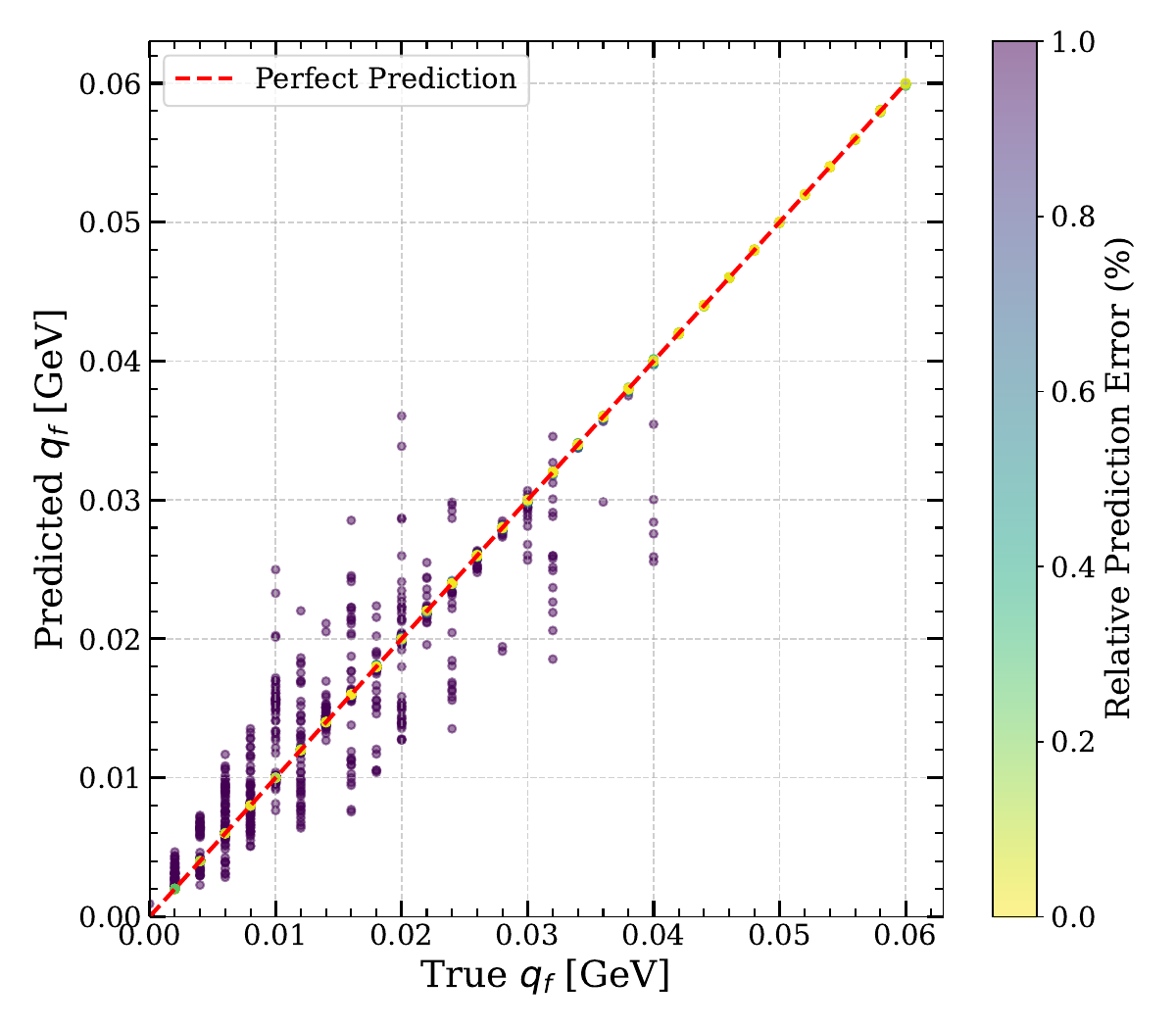}
  \end{minipage}
  \caption{Scatter plots comparing the predicted and true values of the axion-like particle mass $m_\chi$ (left) and dark matter fraction $q_f$ (right) obtained from the machine learning model. The dashed red line represents perfect prediction. Color bars indicate the relative prediction error in percentage.}
  \label{ML_1}
\end{figure*}

\begin{table}[h!]
\centering
\caption{Performance metrics for the machine learning models predicting ALP DM mass ($m_\chi$) and DM fermi momentum ($q_f$)}
\begin{tabular}{lcc}
\hline
\textbf{Metric} & \textbf{$m_\chi$} & \textbf{$q_f$} \\
\hline
\textbf{R$^2$} & 0.9988 & 0.9992 \\
MAE & 3.1661 & \;\; $3.73 \times 10^{-5}$ \\
RMSE & 10.1552 & \;\; $3.99 \times 10^{-4}$ \\
MAPE (\%) & 2.0339 & 0.3399 \\
Aggregate Relative Error \; & 0.00726 & 0.00164 \\
\hline
\end{tabular}
\label{tab:model_metrics}
\end{table}

To provide a comprehensive assessment of the models' performance, we utilized a suite of standard regression metrics. To report the results in their original, physically meaningful units, the model's predictions (which were on the log scale) were converted back to the original scale using an exponential transformation before any metrics were calculated. This rigorous separation of training and testing data ensures that our reported performance metrics are an honest and reliable estimate of the models' ability to generalize to new, unseen data. The primary metric for overall model fit is the Coefficient of Determination (\textbf{R$^2$}), which measures the proportion of the variance in the target variable that is predictable from the independent variables. An \textbf{R$^2$} score of 1.0 indicates a perfect fit. To quantify the prediction error in absolute terms, we calculated the Mean Absolute Error (MAE), which represents the average absolute difference between the predicted and true values, and the Root Mean Squared Error (RMSE), which is more sensitive to large errors. Finally, to assess the performance in terms of percentage error, which was a key goal of our log-transform strategy, we calculated the Aggregate Relative Error. This metric is defined as the sum of all absolute errors divided by the sum of all true values, providing a stable measure of the total relative error across the entire test set. Data related to metrics for both target variables can be found in Table~\ref{tab:model_metrics}.

\begin{figure*}[htbp]
  \centering
  \begin{minipage}[b]{0.47\linewidth}
    \centering
    \includegraphics[width=\linewidth]{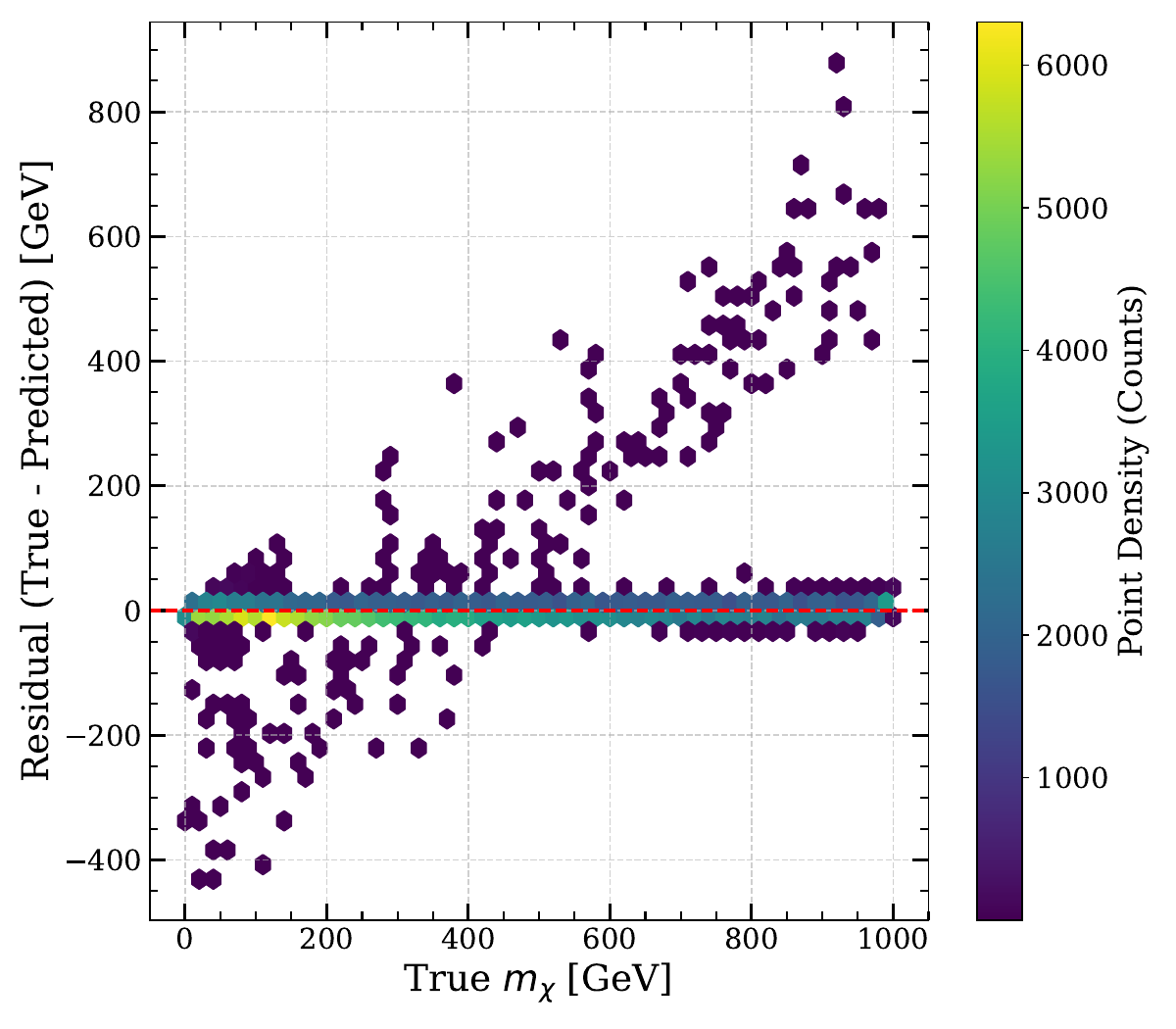}
  \end{minipage}\hspace{0.02\linewidth}%
  \begin{minipage}[b]{0.47\linewidth}
    \centering
    \includegraphics[width=\linewidth]{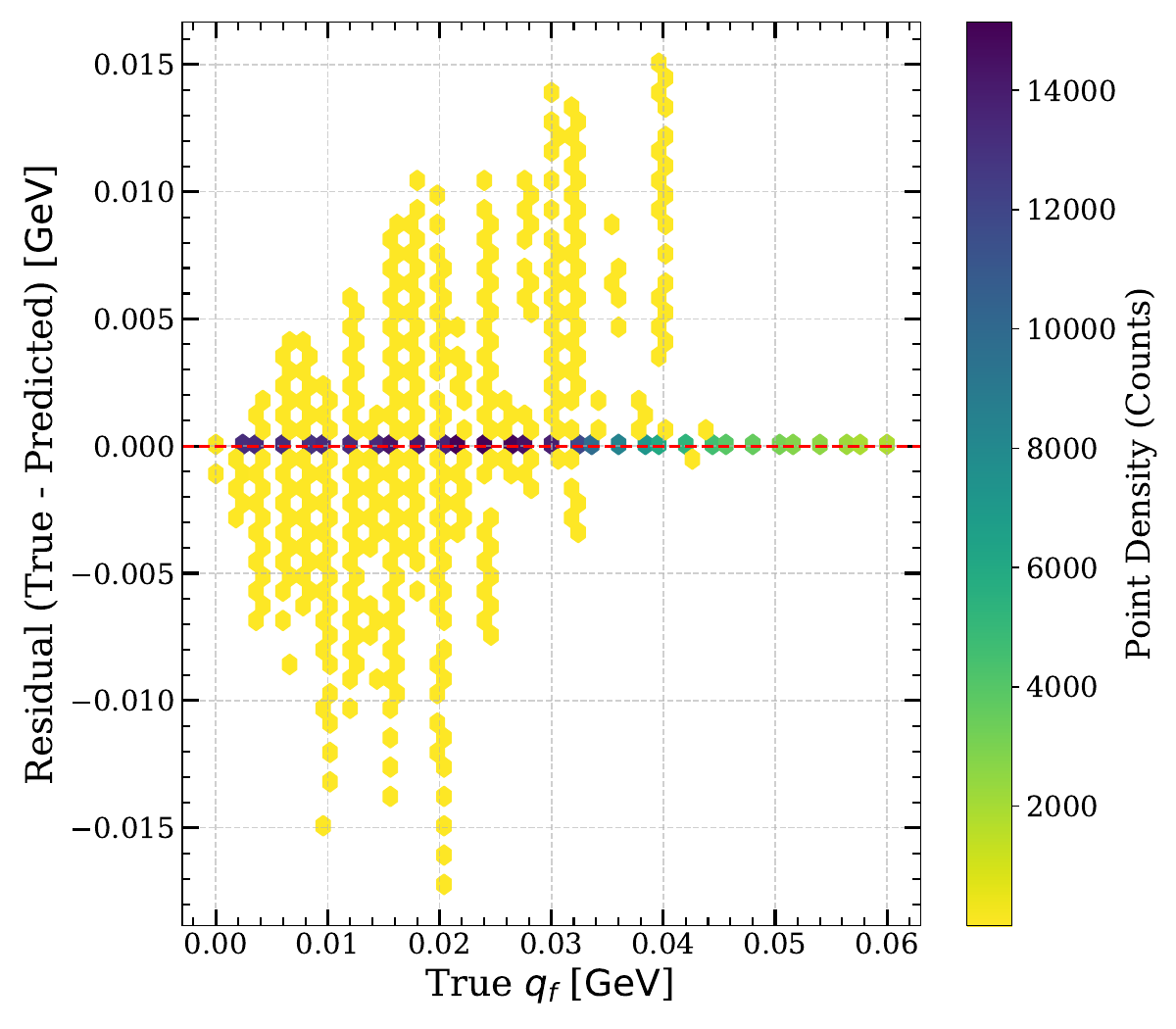}
  \end{minipage}
  \caption{Hexbin density plots of the residuals, defined as $\mathrm{Residual} = \mathrm{True} - \mathrm{Predicted}$, versus the true axion-like particle mass $m_\chi$ (left) and fermi momentum $q_f$ (right) on the unseen hold-out data. The color indicates the number of data points per bin.}
  \label{ML_2}
\end{figure*}

Limitation $\&$ Scope: Our random row-wise split defines the task as supervised interpolation within the densely sampled parameter space of our simulations. Because individual points along the same EoS trajectory are randomly divided between the training and test sets, the evaluation measures the model’s ability to predict unseen configurations that lie within the physical manifold covered by the training grid. This strategy does not test extrapolation to entirely new EoS families or parameter combinations outside this domain. Such extrapolation is inherently ill-posed, since different EoS and DM parameters can yield nearly identical macroscopic $(M, R, \Lambda)$ curves. Accordingly, our results quantify the model’s precision within the simulated domain, rather than its ability to infer physics beyond it.

Figure~\ref{ML_1} presents a comprehensive validation of the final universal regression models on the unseen hold-out test set, providing both visual and quantitative evidence of their high fidelity. The left plot, which compares the predicted and true values of DM mass ($m_\chi$), shows an exceptional level of accuracy. The tight clustering of points along the Perfect Prediction line corresponds to an \textbf{R$^2$} score of 0.9988 and a low mean absolute error (MAE) of just 3.17~GeV. The color scale, representing the relative prediction error, reveals that most predictions (dark purple points) exhibit errors close to zero, in line with the extremely low aggregate relative error of 0.73\%. The few light yellow points represent rare outliers, where the relative error is highest. These are mostly concentrated at $m_\chi < 200$~GeV, where the model occasionally overestimates the mass. This indicates a specific physical regime where the model exhibits higher uncertainty, a valuable insight for future investigation. Nevertheless, the rarity of these errors ensures that the overall performance remains excellent. The right plot illustrates the model’s predictions for the dark matter Fermi momentum ($q_f$), which are even more precise. Here, the model achieves an \textbf{R$^2$} score of 0.9992 and a remarkably low MAE of $3.7\times10^{-5}$. The relative error remains consistently low across the entire range of $q_f$, with an aggregate relative error of just 0.16\%. These results confirm that both regression models are accurate, robust, and well-calibrated, capturing the underlying physical relationships with outstanding precision.

Figure~\ref {ML_2} provides a rigorous validation of the final models by visualizing the distribution of prediction errors (residuals) on the unseen hold-out test set. The left plot, corresponding to the $m_{\chi}$, uses a hexbin plot where color indicates the density of data points. The plot is dominated by a bright yellow/green horizontal line of hexagons centered perfectly on a residual of zero, with the colorbar indicating that these bins contain over 6000 data points each. This demonstrates that the vast majority of predictions are extremely precise. The flat, horizontal nature of this dense region across the entire range of true $m_{\chi}$ values confirms that the model is robust and free of systematic bias. The scattered, darker purple hexagons represent the model's few, infrequent, larger errors, highlighting that significant deviations are rare. The right plot displays the same analysis for the $q_f$. Here, the performance is even more striking; the high-density core (dark purple, with over 14,000 counts per bin) is exceptionally tight around the zero-error line, visually confirming the model's sub-percent relative error and minuscule MAE. The few low-density outliers (yellow points) reveal a subtle structure, suggesting specific physical regimes that are slightly harder to predict, which provides a valuable direction for future investigation. Collectively, both panels demonstrate that the models are not only highly accurate but also well-calibrated and scientifically robust.

\begin{figure*}[htbp]
  \centering
  \begin{minipage}[b]{0.47\linewidth}
    \centering
    \includegraphics[width=\linewidth]{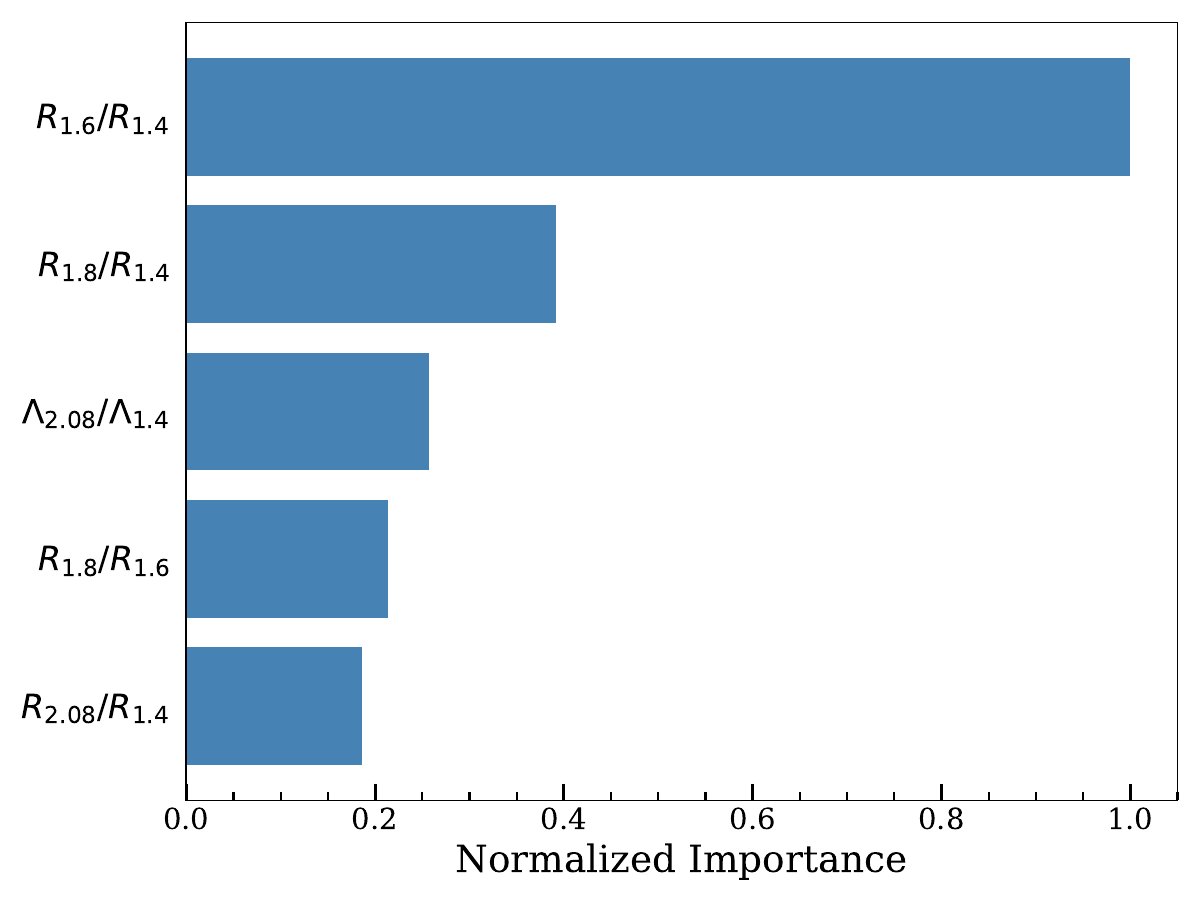}
  \end{minipage}\hspace{0.02\linewidth}%
  \begin{minipage}[b]{0.47\linewidth}
    \centering
    \includegraphics[width=\linewidth]{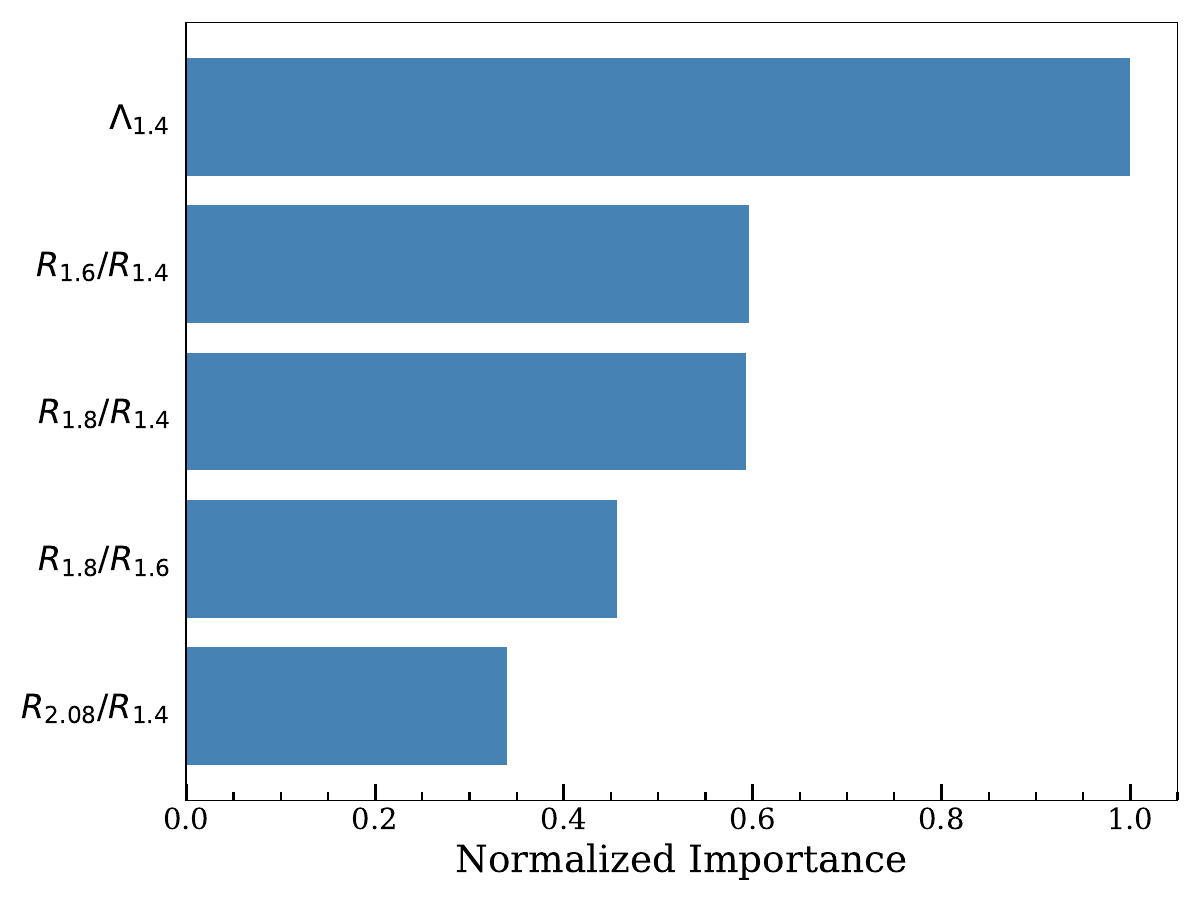}
  \end{minipage}
  \caption{Normalized feature importance for the top five predictors of the axion-like particle mass ($m_{\chi}$, left plot) and fermi momentum ($q_f$, right plot).}
  \label{ML_3}
\end{figure*}

Figure~\ref{ML_3} provides a crucial insight into the physical drivers learned by the models by comparing the normalized feature importances for predicting ($m_{\chi}$, left plot) and ($q_f$, right plot). The x-axis in both plots represents a Normalized Importance scale, where the single most influential feature for each model is assigned a score of 1.0. All other features are then scored relative to this most important one, allowing for a clear, direct comparison of their relative impact. The left plot reveals that the prediction of $m_{\chi}$ is overwhelmingly dominated by global shape indicators rather than local observables. The most important feature, with a score of 1.0, is the ratio of the radius at 1.6 to 1.4 solar masses ($R_{1.6}/R_{1.4}$). This indicates that the model found the ``stiffness" or the rate of change of the radius with increasing mass to be the most critical predictive information. These findings are broadly consistent with the earlier study by \citet{Thakur:2024mxs}, where radius measurements, particularly at the extremes of the mass range, were identified as key features. The other top features are also ratios of radii or tidal deformabilities, confirming that the model has learned a relationship based on the overall structural behavior of the star. 
In contrast, the right plot demonstrates that the model has learned a different physical dependence for $q_f$. While its prediction is most strongly anchored by a single key parameter: the tidal deformability at a canonical 1.4 solar masses ($\Lambda_{1.4}$), which has a normalized importance of 1.0. But the evolutionary radius ratios (e.g., $R_{1.6}/R_{1.4}$) are still highly relevant, ranking second and third with importances of approximately 0.6. This suggests that while $q_f$ leaves a distinct local imprint potentially observable in single events, its inference is significantly refined by the contextual information provided by the population-level analysis. Collectively, these results confirm that a robust simultaneous extraction of $(m_{\chi}, q_f)$ requires the multi-point structural information necessitating data from an ensemble of neutron star observations. In a realistic observational scenario, this implies a two-step inference pipeline. First, the continuous MR relation must be reconstructed from the ensemble of discrete measurements (e.g., NICER posteriors and LIGO tidal constraints) using standard Bayesian techniques such as Gaussian processes~\cite{Landry:2020vaw} or parametric methods~\cite{Raaijmakers:2019dks}. Second, the multi-point structural features required by our model (e.g. $R_{1.6}$/$R_{1.4}$) are extracted from this reconstructed curve. This workflow allows our machine learning framework to map the global geometry of the EoS back to the dark matter parameters.

\section{Discussion $\&$ Summary}\label{discus&summary}

This study employs Neutron Stars (NSs) as astrophysical laboratories to probe axion-like particle (ALP) mediated dark matter (DM). We implement the DM sector within a realistic non-linear RMF framework, considering both stiff and soft hadronic equations of state (EoSs) consistently extended with a BPS crust.  A comprehensive grid scan across the DM parameter space, covering the mass $m_\chi$ and Fermi momentum $q_f$, yielded more than $30{,}000$ EoSs, for each NS model. These were subsequently filtered against multi-messenger observational constraints to isolate the physically viable regions. The central result is that the inclusion of DM systematically softens the EoS, which in turn modifies the global stellar structure. This softening manifests as a predictable downward shift in the mass--radius relation: increasing either $m_\chi$ or $q_f$ leads to smaller radii and, most importantly, a reduced maximum mass $M_{\rm max}$.

The outcome of these findings is shown in Figure~\ref{mr}, which presents the final allowed regions in the DM parameter space, with stiff as well as soft hadronic EoSs, after applying the joint constraints of $M_{\mathrm{max}} \geq 2.0\,M_\odot$ and $\Lambda_{1.4} \leq 580$. A clear contrast emerges between the two nuclear models. For the stiff EoS1, the constraints define a stable ``island'' bounded by the tidal deformability and maximum mass limits, suggesting that a non-zero DM component may even be favored. In contrast, the soft EoS18 leaves only a narrow strip at low $q_f$, as the maximum mass bound excludes most of the space. This highlights the strong discriminatory power of multi-messenger observations and shows that improved knowledge of the nuclear EoS will directly sharpen DM constraints. Importantly, the ALP-mediated DM model provides a unified explanation across the mass spectrum, from massive pulsars like PSR~J0740+6620 to compact objects such as HESS~J1731--347, minimizing the need for separate theoretical models for different neutron star populations. The probabilistic analysis shows that the DM Fermi momentum is relatively well-constrained, with $q_f = 0.034^{+0.020}_{-0.012}$, while the DM particle mass remains broad, spanning 101--949\,GeV (90\% CI). This indicates that NS observations are more sensitive to the overall DM content than to the particle mass, highlighting a model degeneracy that requires further study. 

To capture the nonlinear links between model parameters and NS observables, we combined statistical and machine learning methods. While Pearson correlations suggested a weak role for $m_\chi$, Mutual Information revealed stronger nonlinear effects, motivating a broader framework. SHAP analysis of XGBoost models clarified the hierarchy: the nuclear EoS type governs global properties like $M_{\rm max}$, $q_f$ dominates $\Lambda$ and $R_{1.4}$, while $m_\chi$ plays a secondary but consistent role. Tidal deformability thus emerges as a key probe of DM in NSs. We further developed high-precision regression models to infer $m_\chi$ and $q_f$ directly from observables. With $R^2 > 0.998$ and negligible relative errors, the models capture the complex mapping from NS data to DM parameters, offering a fast alternative to grid scans. Feature analysis shows that $m_\chi$ is best constrained by radius ratios (e.g., $R_{1.6}/R_{1.4}$), while $q_f$ is tightly linked to $\Lambda_{1.4}$. This provides clear guidance for observations: multi-mass radius measurements constrain the DM particle mass $m_\chi$, whereas precise $\Lambda_{1.4}$ is key for the DM Fermi momentum $q_f$. Together, these findings sharpen strategies for future multi-messenger probes of DM in neutron stars.

This distinct mapping provides a clear observational strategy for a potential 'smoking gun': a set of precise measurements (e.g., of $\Lambda_{1.4}$ and $R_{1.6}/R_{1.4}$) that are \textit{mutually contradictory} for any single baryonic EoS, which rigidly links these properties, could be a compelling signature of the extra degrees of freedom our DM-admixed model provides.

In conclusion, this study provides stringent constraints on ALP-mediated DM and reveals its intricate dependence on the underlying nuclear physics, highlighting the nontrivial behaviour of DM in NS matter. Moreover, the integration of statistical methods with machine learning establishes a powerful framework for probing DM physics in dense matter and opens promising directions for future multi-messenger astrophysics. In future work, we aim to extend our analysis by incorporating additional exotic degrees of freedom, such as hyperons and quarks \cite{Gholami:2024ety, Christian:2025dhe, Thakur:2025qwl, Rather:2021yxo}, into the nuclear matter sector to obtain a more complete description of dense matter interactions and study the oscillation modes with these different compositions of the hadronic matter \cite{Rather:2023dom, Pradhan:2022vdf, Rather:2024hmo, Rather:2023tly, Shirke:2024ymc, Thakur:2025zhi}. An additional avenue for future work is to extend the machine learning framework into a more complete inference approach, allowing us not only to highlight key observables but also to provide quantitative constraints on DM parameters as observational data become increasingly precise.


\section*{Acknowledgement}
 
IAR acknowledges support from the Alexander von Humboldt Foundation. P. Thakur is supported by the National Research Foundation of Korea (NRF) grant funded by the Korea government (MSIT) (No. RS-2024-00457037). This work was supported (in part) by the Yonsei University Research Fund(Yonsei University Frontier Fellowship for Postdoctoral Researchers) of 2025. This research is supported by the National Research Council of Thailand (NRCT) : (Contact No. N41A670401). CP has received funding support from the NSRF via the Program Management Unit for Human Resources \& Institutional Development, Research and Innovation [grant number B39G680009]. CP is supported by the Fundamental Fund 2569 of Khon Kaen University.

\section{Appendix}
\label{appendix}

\begin{figure*}[htbp!]
  \centering
  \begin{minipage}[b]{0.47\linewidth}
    \centering
    \includegraphics[width=\linewidth]{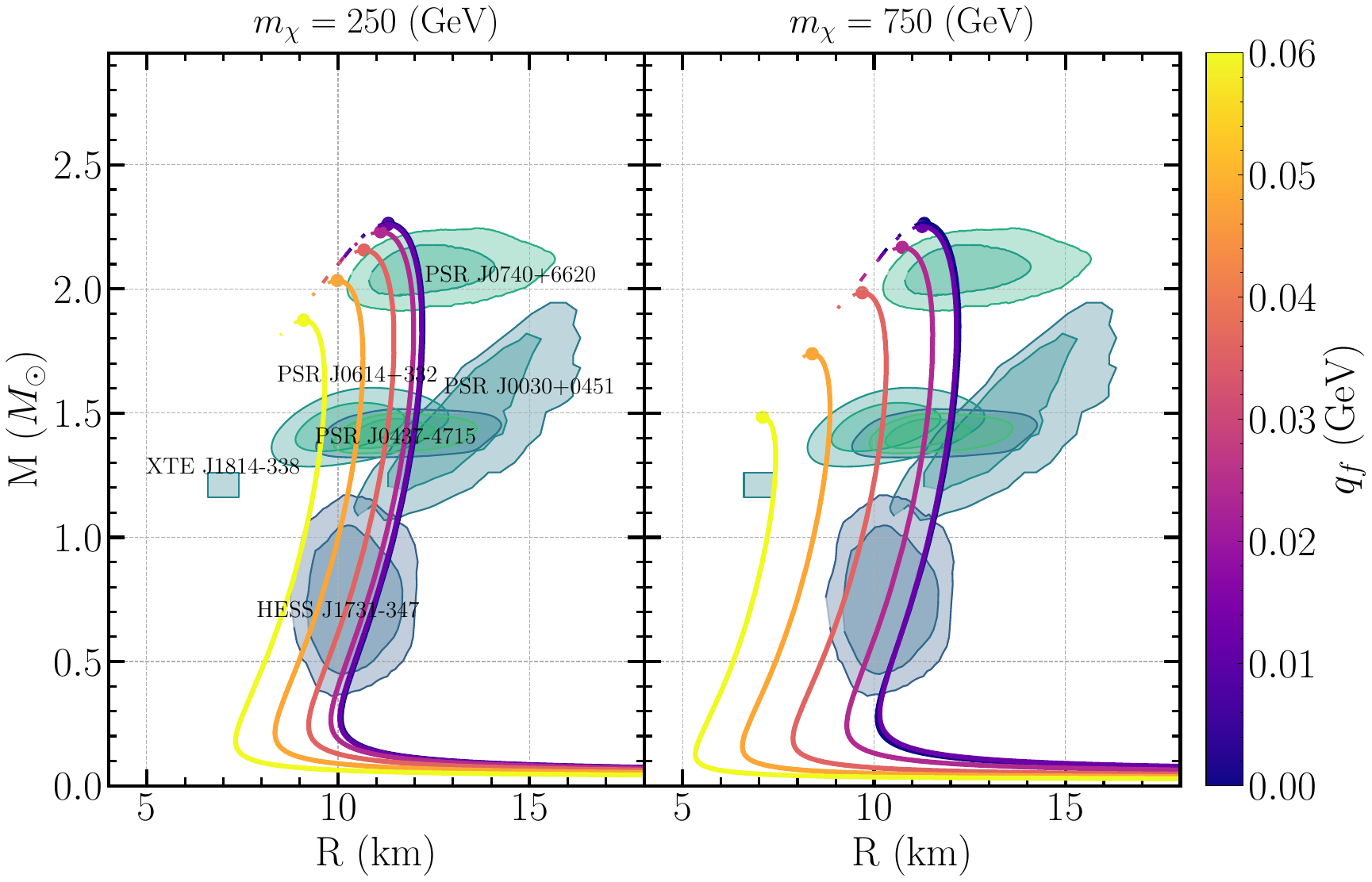}
  \end{minipage}\hspace{0.02\linewidth}%
  \begin{minipage}[b]{0.47\linewidth}
    \centering
    \includegraphics[width=\linewidth]{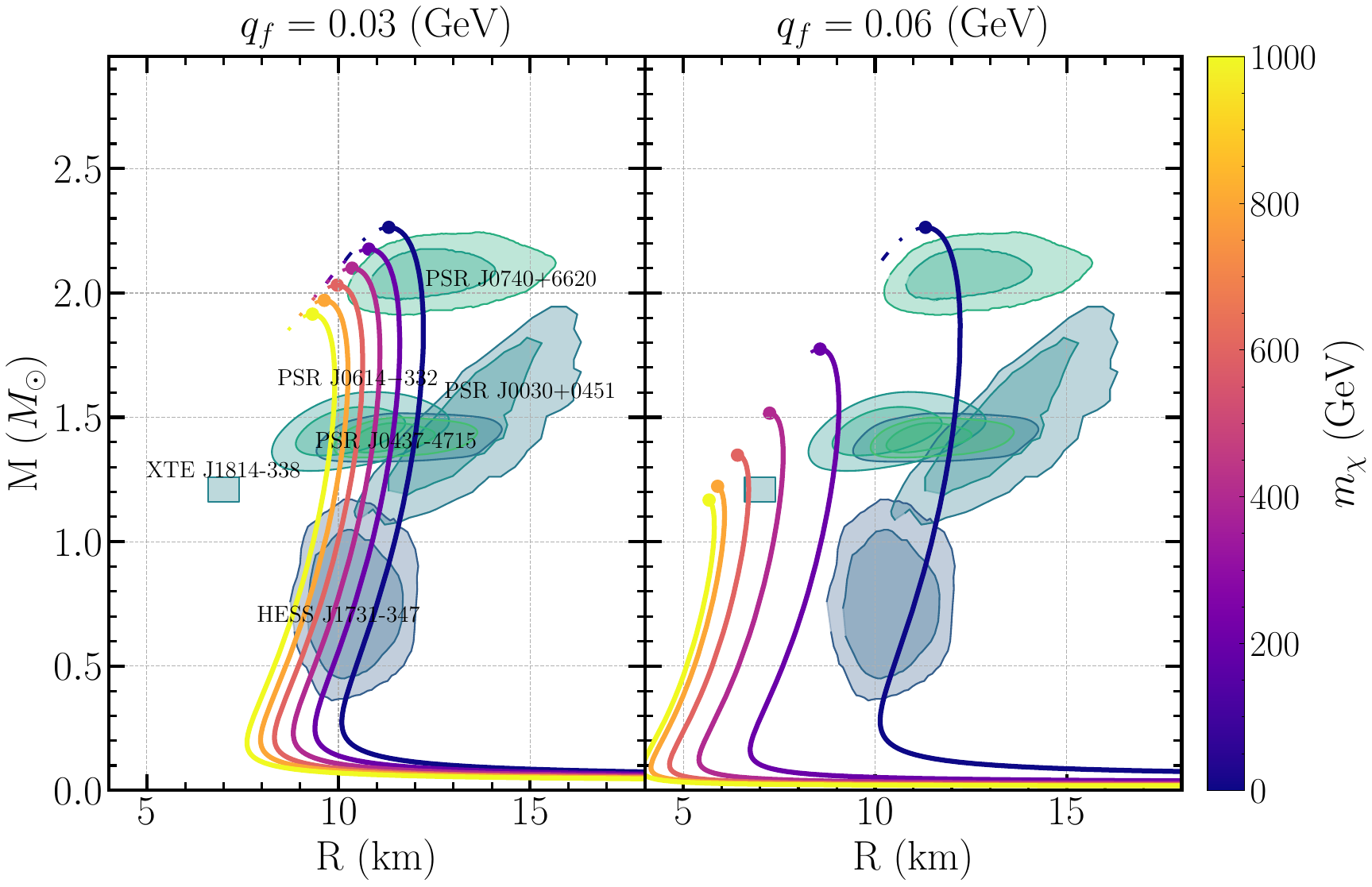}
  \end{minipage}
   \caption{Same as \ref{MR_fix}, but with the soft hadronic EoS, EoS18.}
\label{MR_fix_soft}
\end{figure*}

Here we show some results that were discussed and compared to those in the discussion part.

Figure~\ref{MR_fix_soft} illustrates the MR relations for the soft hadronic model, EoS18, incorporating ALP-mediated dark matter. The inclusion of DM systematically softens the composite EoS, leading to a pronounced reduction in both the maximum mass ($M_{\text{max}}$) and the stellar radius for any given mass. This effect intensifies with a larger DM fraction (increasing $q_f$) and is more severe for lower DM particle masses ($m_\chi$). Due to the inherent softness of the EoS18, even a modest DM admixture is sufficient to suppress $M_{\text{max}}$ below the crucial $2\,M_\odot$ observational limit, thereby rendering a large portion of the parameter space unviable.

When compared with the stiff hadronic EoS1 results as shown in Figure \ref{MR_fix},  the stiff EoS provides a significantly higher baseline $M_{\text{max}}$, creating a substantial buffer against the softening induced by the DM component. Consequently, the stiff model remains compatible with the $2\,M_\odot$ constraint over a much broader range of DM parameters. While certain configurations with high DM content in the soft EoS model can produce highly compact stars consistent with objects like XTE J1814-338, these solutions invariably violate the maximum mass constraint, highlighting a critical tension that is far less pronounced in the more resilient stiff EoS framework.

\bibliographystyle{apsrev4-112}
\bibliography{references} 

\end{document}